\documentclass{nature} 
\usepackage{graphicx}
\usepackage{color,soul}
\usepackage{physics}
\newcommand{\tsb}[1]{\textsubscript{#1}}
\newcommand{\tsp}[1]{\textsuperscript{#1}}
\usepackage[normalem]{ulem}

\title{Downfolding the Molecular Hamiltonian Matrix using Quantum Community Detection}

\author{Susan M. Mniszewski$^1*$, Pavel A. Dub$^2$, Sergei Tretiak$^3$, Petr M. Anisimov$^4$, Yu Zhang$^3$, \& Christian F. A. Negre$^3$}

\begin{document}

\maketitle

\begin{affiliations}
 \item Computer, Computational and Statistical Sciences Division, Los Alamos National Laboratory, Los Alamos, NM
 \item Chemistry Division, Los Alamos National Laboratory, Los Alamos, NM
 \item Theoretical Division, Los Alamos National Laboratory, Los Alamos, NM
 \item Accelerator Operations and Technology Division, Los Alamos National Laboratory, Los Alamos, NM
\end{affiliations}

\begin{abstract}
Calculating the ground state energy of a molecule efficiently is of great interest in quantum chemistry. The exact numerical solution of the electronic Schrödinger equation remains unfeasible for most molecules 
requiring approximate methods at best. In this paper we introduce the use of \emph{Quantum Community Detection} performed using the D-Wave quantum annealer to reduce the molecular Hamiltonian matrix without chemical knowledge. Given a molecule represented by a matrix of Slater determinants, the connectivity between Slater determinants is viewed as a graph adjacency matrix for determining multiple communities based on modularity maximization. The resulting lowest energy cluster of Slater determinants is used to calculate the ground state energy within chemical accuracy. The details of this method are described along with demonstrating its performance across multiple molecules of interest and a bond dissociation example. This approach is general and can be used as part of electronic structure calculations to reduce the computation required. 
\end{abstract}


\section*{Introduction}
At this stage in quantum computing, it is useful to demonstrate the application of quantum algorithms to real-world problems even at small scale where classical solutions are available, though may be approximate. This allows for verification and validation, and helps us prepare for the larger-scale problems with unknown results on future quantum hardware. Noisy Intermediate-Scale Quantum (NISQ) technology has provided us the opportunity to explore new approaches 
to simulate chemistry and physics~\cite{Preskill2018-wy}. 
By taking advantage of quantum-mechanical effects, NISQ devices 
promise to improve solutions, provide new pathways to solutions and even solve the problems that are intractable on current classical computers. 
In addition, quantum formulations of complex network algorithms have gained interest in quantum information science~\cite{Biamonte2019}. The cross-disciplinary approach of combining network science and quantum computing has provided new ways to solve chemistry and physics problems that can be framed as graphs. In this work we explore downfolding or reducing the Hamiltonian matrix for electronic structure calculations using \emph{Quantum Community Detection}\cite{Ushijima-Mwesigwa2017,Negre2020}  on the D-Wave Quantum Annealer. 
The approach does not rely on any prior knowledge about the chemical system of interest and is based on a matrix formulation only.


Computational chemistry aims to solve the Schrödinger equation numerically for many-body quantum systems of interest such as molecules or solid-state materials. For the vast majority of applications, we are interested in solving the time-independent, non-relativistic form of this equation for electrons by treating nuclei as classical point charges (using the Born-Oppenheimer approximation). The resulting 
electronic Hamiltonian can be solved exactly within the space spanned by the finite one-electron orbital basis set or a numerical grid. The corresponding method is known as full configuration interaction (FCI) or the diagonalization of the Hamiltonian matrix in the basis of Slater determinants (SDs). SDs form a complete basis of the problem in Hilbert space accounting for the exchange and correlation energy of the multielectronic system. The FCI method provides an unambiguous standard with which to compare more approximate methods. Since FCI scales factorially with the number of electrons and spin-orbitals, its practical implementation is limited to very small molecules/basis sets. For example, molecular dicarbon (C\tsb{2}) in aug-cc-pVTZ basis (with 12 electrons to be distributed on 184 spin-orbitals) is one of the largest molecules ever simulated by means of FCI using the Oak Ridge National Lab Cray-X1 supercomputer \cite{Gan2005}. To overcome these limitations, various approximate solutions have been developed by either truncating the problem in a reduced user-defined active space (e.g. multiconfigurational self-consistent methods such as the complete active space self-consistent field (SCF) method (CASSCF) \cite{Roos1980,Siegbahn1980,Siegbahn1981}) or in a few SDs by employing Configurational Interaction (CI) or Coupled Cluster (CC) approximations \cite{Szabo1996}. Examples of the truncation methods include CI + single excitations (CIS), CI + single and double excitations (CISD), CI + single, double, and triple excitations (CISDT), CI + single, double, triple, and quadruple excitations (CISDTQ), and coupled cluster single-double-triple (CCSDT), to name a few. In their appreciable level of flavor, they scale as ~$\mathcal{O}(N^{7-9})$ and are also applicable only to few atom systems. The Hartree-Fock (HF) mean-field approximation reduces the problem to a single electron moving in an average field of others with ~$\mathcal{O}(N^3)$ complexity, and serves as an upper bound for an energy representing an uncorrelated system of electrons.


Quantum chemistry is regarded as one of the first disciplines that will benefit from quantum computing. Predicting the properties of atoms and molecules will become one of the most important applications of quantum computers \cite{Feynman1982,McArdle2020,Cao2019}.
Specifically, quantum computing promises to revolutionize quantum simulations of materials by bringing down the intractable exponential cost on classical computers [~$\mathcal{O}(\alpha^N)$] to a polynomial scaling on quantum computers.
This can be achieved by using either of the two available forms of quantum computation: gate-based quantum computing and adiabatic quantum computing.
Current quantum computers are hardware-limited in both the number and quality of qubits, qubit connectivity, presence of high noise levels, and the need for full error-correction \cite{Shaydulin2019}. Noisy gate-based quantum computers are currently available at the scale of \textasciitilde50-70 qubits 
Adiabatic quantum computers are available in the form of quantum annealers, such as the D-Wave 2000Q (with \textasciitilde2048 qubits) and the upcoming D-Wave Advantage machine (with \textasciitilde5000 qubits). 

Solutions to electronic structure problems have been demonstrated for small molecules on gate-based quantum computers \cite{Kandala2017,Nam2019,Omalley2016,Aspuru-Guzik2005} and the D-Wave 2000Q quantum annealer \cite{Genin2019,Xia2018,Streif2019}. Polynomial scaling has not yet been demonstrated for the electronic structure problem on a quantum annealer.
However, as we show in this work, the most advanced D-Wave 2000Q quantum annealer is well suited to solve combinatorial optimization problems, which can be used to reduce the complexity of the electronic structure problem. 
%
In this work we reduce the molecular Hamiltonian on a quantum annealer and then calculate the approximate ground and/or excited state energies on a classical computer through diagonalization. 
The advantages of solving the graph partitioning and community detection problem on a Quantum Annealer were recently demonstrated in \cite{Ushijima-Mwesigwa2017,Negre2020}. 
The smaller sub-matrices resulting from the clusters of SDs are then candidates for the best low energy solution. The one containing the lowest energy SD (\emph{i.e.} the HF SD) or resulting in the lowest gauge metric (described later) is the cluster that provides the best approximation of the original system. The energy is calculated through diagonalization on a classical computer. This typically results in energy many times within desired chemical accuracy ($\leq$ 1.6e-03 Hartrees or 1 kcal/mol).

A quantum annealer, such as the D-Wave 2000Q \cite{DWave2000Q}, is able to solve graph problems as combinatorial optimization framed as an Ising model or quadratic unconstrained binary optimization (QUBO) problem. Quantum annealing uses the quantum-mechanical effects of tunneling, superposition and entanglement~\cite{McGeoch2014-sj} to minimize and sample from energy-based models. Evidence that these effects play a useful role in the processing have been discussed in \cite{Albash2015,Boxio2016,Lanting2014}. The following objective function in Ising Model form is minimized.

\begin{equation}
    O(\textbf{h,J,s}) = \sum\limits_{i}h_{i}s_i + \sum\limits_{i<j}J_{ij}s_is_j
    \label{eq:dwave}
\end{equation} 
where $s_{i} \in$ $\{-1,+1\}$ encodes the binary results; 
$h_{i}$ and $J_{ij}$ represent the weights on the qubits and strengths on the couplers between qubits of the problem Hamiltonian. 
The D-Wave quantum computer is composed of qubits with sparse connectivity as a fixed sparse graph, known as a \emph{Chimera} graph. 
Qubits are in a ``superposition'' state (both a ``-1'' and a ``+1'' simultaneously) during the annealing process. Once the annealing is done, each qubit 
settles into an Ising spin value $\in$ ${-1,+1}$, resulting in a low-energy ground state. Due to sparse connectivity, a logical variable maps to a chain of qubits \cite{Lanting2014} in the D-Wave quantum processing unit (QPU).

Quantum annealers are useful for tackling NP-hard complex problems including optimization, machine learning and sampling.
Maximization can also be solved by using the negative of Eq.~\ref{eq:dwave} as the objective function.
The QUBO formulation where variables $x_i$ take values of either $0$ or $1$ is an alternative representation.
The QUBO and Ising Model are related by the following linear transformation: $s = 2x - 1$. 
  
\emph{Quantum Community Detection} is Community Detection \cite{Newman2006,Newman2004} formulated as a QUBO problem and is run on the D-Wave quantum annealer. 
The community detection algorithm is an unsupervised machine learning technique that allows for the discovery of network substructure as tightly knit clusters or communities. The quantum version of this algorithm as a QUBO problem running on a quantum annealer uses superposition to explore the design space and settles into a low-energy solution.
The molecule Hamiltonian is treated as an adjacency matrix for a graph with weighted edges. The modularity matrix $B$ is constructed from the adjacency matrix $A$ and degree of each node (number of edges connected to a node), $g_i$, 
where, $2m = \sum_i g_i$.

\begin{equation}
    B_{ij} =  A_{ij} - \frac{g_ig_j}{2m} =  A_{ij} - \frac{g_ig_j}{\sum_{i}g_{i}}
    \label{eq:Qij}
\end{equation}

The objective function for the optimal modularity $Q$ as a QUBO is maximized as shown in Eq.~\ref{eq:mod_obj}. The elements of $x$ are the solution and are binary, where $x_i$ is $0$ or $1$. Details of the mathematical formulation for multiple communities have been developed previously and are available~\cite{Negre2020}. When solving for multiple communities, a \emph{one hot encoding} or super-node concept is used with unary encoding. The objective function is shown as the following.

\begin{equation}
    Q(x) = max(x^TBx)
    \label{eq:mod_obj}
\end{equation}

The D-Wave 2000Q is limited in the number of variables that can be embedded and solved directly on the hardware. Therefore, the largest fully connected graph that can be embedded is 65. A hybrid quantum-classical approach is required to address larger numbers of variables as nodes (here representing SDs)~\cite{Shaydulin2019}. We use the D-Wave developed \emph{qbsolv} classical solver \cite{Booth2017} to orchestrate the QUBO solution process between the central processing unit (CPU) and QPU. Global minimization is performed by \emph{qbsolv}, making multiple calls to the D-Wave to solve sub-QUBOs. This is followed by local minimization using tabu search. The D-Wave Ocean application programming interface (API) or the command line is used to initiate this process. A low-energy solution is returned as a bitstring of zeros and ones that requires translation based on the problem representation.

\begin{figure}
\includegraphics[width=0.90\linewidth]{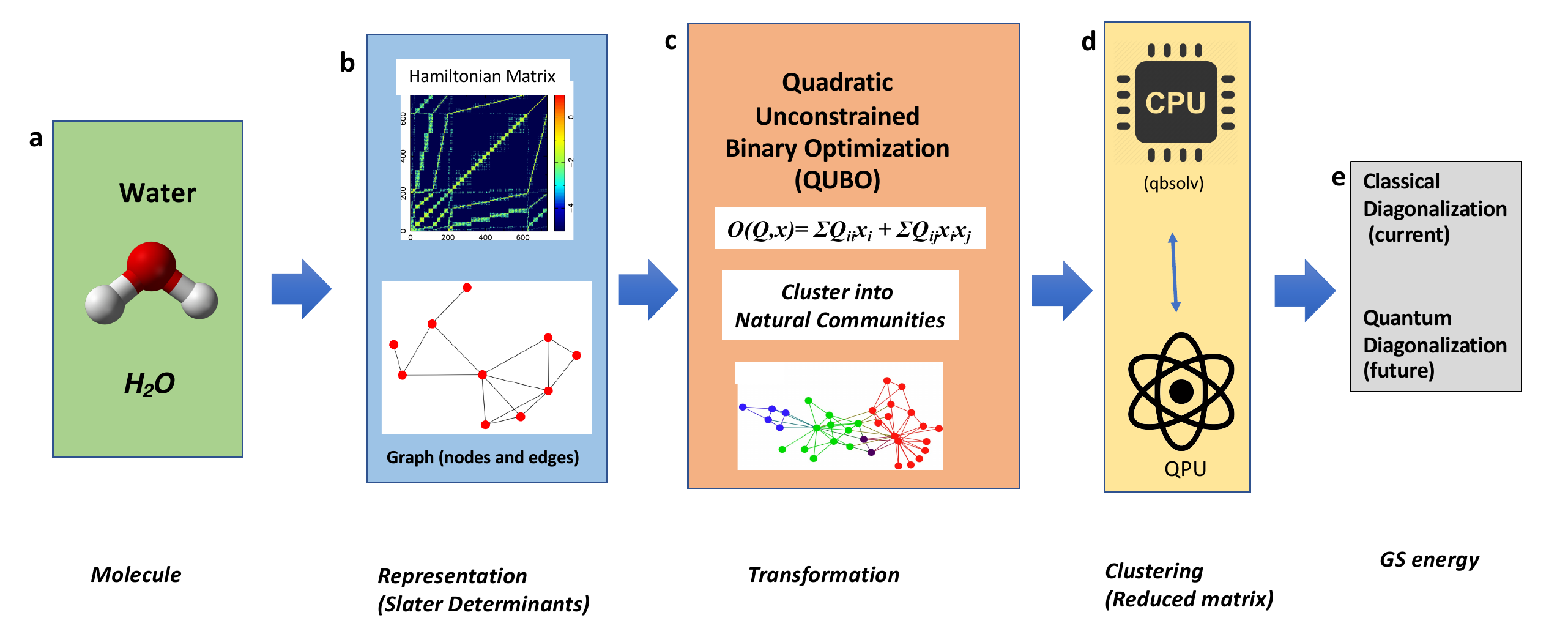}
\caption{Conceptual workflow for the downfolding process. \textbf{a} Starting from Cartesian co-ordinates  for  atoms,  atomic  numbers  and  a  chosen  basis  set,  an  electronic  Hamiltonian  matrix is constructed in the basis of Slater Determinants (SDs) of spin-orbitals by using a classical quantum chemistry code. \textbf{b} This matrix is further represented as a graph where each SD is a node.  \textbf{c} Transformation to a Quantum Unconstrained Binary Optimization (QUBO) form of \emph{Community Detection} is run using quantum-classical \emph{qbsolv} + D-Wave. \textbf{d} The resulting clusters/communities as sub-matrices are candidates for the best low energy solution. \textbf{e} The cluster containing the HF SD (or identified using the gauge metric) is the cluster of choice and the ground and/or excited state energy is calculated using diagonalization.}
\label{fig:workflow}
\end{figure}

The conceptual workflow for this process is shown in Fig.~\ref{fig:workflow}. 
Starting from the Cartesian coordinates for atoms, atomic numbers and a chosen basis set, an electronic Hamiltonian matrix is constructed in the basis of SDs of spin-orbitals by using one of the classical quantum chemistry codes such as Psi4 \cite{Parrish2017} (see Fig.~\ref{fig:workflow}\textbf{a}). 
 This matrix as a graph is transformed into a QUBO form of \emph{Community Detection} to be solved on the D-Wave Quantum Annealer (see Fig.~\ref{fig:workflow}\textbf{b}). Due to the matrix/graph size, a quantum-classical approach using \emph{qbsolv} and the D-Wave are required (see Fig.~\ref{fig:workflow}\textbf{c}). The resulting clusters or communities are then candidates for the best low-energy solution (see Fig.~\ref{fig:workflow}\textbf{d}). The one containing the HF SD (or identified from the gauge metric) is the low-energy cluster and the ground state energy is calculated by diagonalization on a classical computer (see Fig.~\ref{fig:workflow}\textbf{e}). 

\section*{Results}

\subsection{The Downfolding Process.}
The downfolding approach proceeds as follows.
A molecular Hamiltonian $\hat{H}$ matrix in the basis of SDs 
serves as input. 
The \emph{k-clustering} sets of the Hamiltonian matrix are explored starting with $k=2$.
\emph{Quantum Community Detection} performed using the D-Wave quantum annealer determines the $k$ communities or clusters. 
Each of the $k$ clusters translates into a sub-matrix of SDs. The one containing the HF SD (or identified by the gauge metric) is the cluster of interest out of the $k$ clusters and yields the lowest energy,
$E_{CL}$, when diagonalized classically.
The results obtained by this procedure fall between the HF ($E_{HF}$) and FCI energy of the full matrix ($E_{FCI}$), with a tendency to be closer to $E_{FCI}$.
The accuracy of the result can be measured as the difference with respect to $E_{FCI}$, $(\Delta = E_{FCI}-E_{CL})$. 
This can be repeated for \emph{3-}, \emph{4-}, \emph{5-clustering} and more as needed. Each \emph{k-clustering} is independent of the others. One does not need to go through the \emph{k-clusterings} in order, but it is natural to do so. Based on knowledge about a molecule, one may even choose to start from a large $k$ and work backwards.
The cluster with the lowest energy delta over all \emph{k-clustering} sets is the best result from an accuracy perspective.
This sub-matrix is the downfolded (or reduced) matrix, constituting an approximate representation of the full Hamiltonian.
The goal of this procedure is to discover a reduced matrix with an energy delta within chemical accuracy. The focus can be on accuracy or size.
One should expect the larger the size, the higher the accuracy.
Multiple clusters with deltas within chemical accuracy may be found. Any of these is sufficient for ground state energy calculations.
We note here that the downfolding method is not limited to the energy and could potentially be extended to compute the atomic forces, for geometry optimization or molecular dynamics. 

\subsection{Illustrating the method with H\tsb{2}O.}
\begin{figure}
\includegraphics[width=0.50\linewidth]{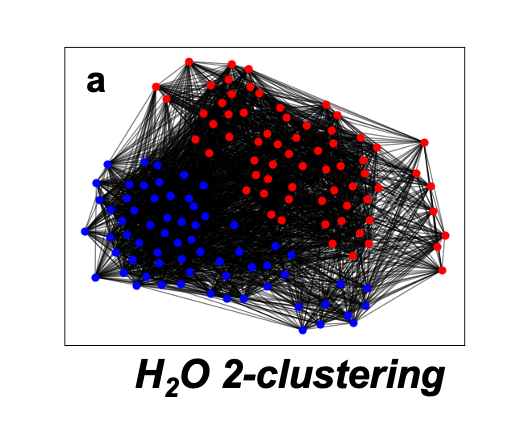}
\includegraphics[width=0.50\linewidth]{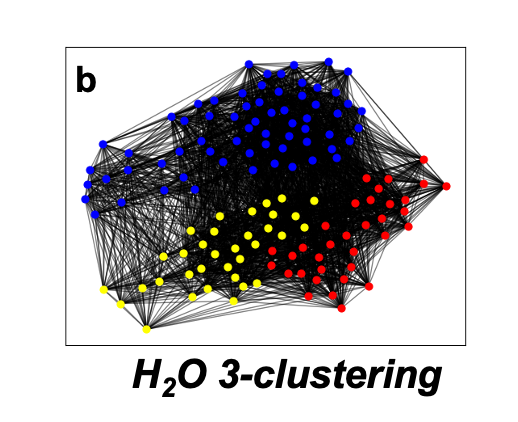} 
\includegraphics[width=0.50\linewidth]{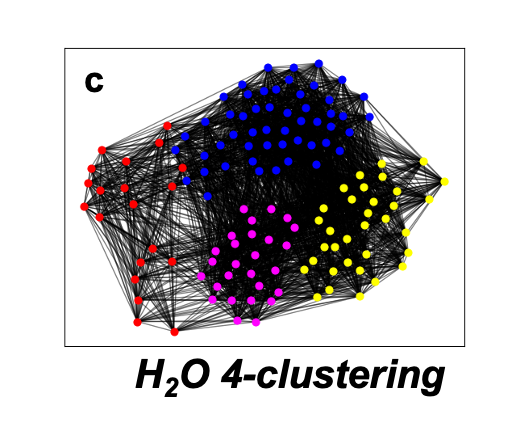}
\includegraphics[width=0.50\linewidth]{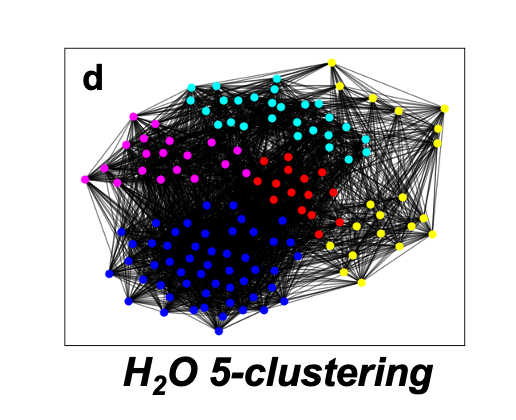} 
\caption{Showing H\tsb{2}O \emph{2-}, \emph{3-}, \emph{4-}, and \emph{5-clustering} from \emph{Quantum Community Detection} on the D-Wave Quantum Annealer. For all the \emph{k-clusterings}, the cluster or community of nodes with the lowest energy is shown in blue. \textbf{a} The \emph{2-clustering} produced communities of size 68 (red) and 65 (blue). \textbf{b} The \emph{3-clustering} is shown with communities of size 34 (red), 65 (blue), and 34 (yellow). The \emph{2-} and \emph{3-clustering} share the same low energy cluster of 65. \textbf{c} The \emph{4-clustering} produced communities of size 20 (red), 52 (blue), 34 (yellow), and 27 (magenta). \textbf{d} The \emph{5-clustering} resulted in communities of size 15 (red), 52 (blue), 20 (yellow), 19 (magenta), and 27 (cyan). The \emph{4-} and \emph{5-clustering} share the same low energy cluster of 52.}
\label{fig:h2o_kcom}
\end{figure}

The Hamiltonian FCI matrix of the H\textsubscript{2}O molecule in the sto-3g minimal basis set is a 133 x 133 sparse Hermitian (i.e., real symmetric) matrix after imposing the unitary groups U(1) and SU(2) (spin and particle conservation) and C$_{2V}$ point group  symmetries. 
The sparse matrix can be represented as a weighted graph of 133 nodes with 3032 edges. Each off-diagonal matrix element is considered an edge weight. Running \emph{Quantum Community Detection}
for \emph{2-clustering} produces the communities shown in Fig.~ \ref{fig:h2o_kcom}\textbf{a}.
The 65 nodes of the blue community represent the SD indices of the reduced matrix at this level with the lowest energy. The energy delta for this cluster is 0.05 kcal/mol for the lowest eigenvalue (ground state energy), which is achieved by almost a 2-fold decrease in matrix size. 
The \emph{3-clustering} shown in Fig.~\ref{fig:h2o_kcom}\textbf{b} produces two other communities and the blue community with the same ground state energy. A further reduction occurs with \emph{4-clustering} as shown in  Fig.~\ref{fig:h2o_kcom}\textbf{c}. Classical diagonalization of this blue community of 52 SDs produces a ground state energy value with an acceptable delta of 0.68 kcal/mol. No further reduction is seen with \emph{5-clustering}, as shown in Fig.~\ref{fig:h2o_kcom}\textbf{d}. A significant 49\% and 39\% reduction in size of the molecule Hamiltonian matrix is realized with the 65 and 52 SD communities. 

\begin{table}
\centering
\caption{Molecules using the FCI method and the sto-3g minimal basis set.}
\medskip
\begin{tabular}{lccccc}
\hline
 & & & Reduced & $\Delta_{FCI-CL}$\\
Molecule & Size & k & Size & (kcal/mol)\\
\hline
\hline
H\tsb{2}O & 133 & 2 & 65 & \hl{0.05}\\
 & & 4 & 52 & \hl{0.68}\\
\hline
CO & 3648 & 2 & 2013 & \hl{0.01}\\
 & & 2 & 1768 & \hl{0.05}\\
 & & 3 & 1768 & \hl{0.05}\\
 & & 4 & 1143 & 4.92\\
 & & 5 & 792 & \hl{0.19}\\
\hline
CH\tsb{4} & 4076 & 2 & 2284 & \hl{0.18}\\
 & & 3 & 1284 & \hl{0.26}\\
\hline
BH\tsb{4}\tsp{-} & 4076 & 2 & 2284 & \hl{0.29}\\
  & & 3 & 1284 & \hl{0.41}\\
\hline
H\tsb{4}O\tsp{2+} & 15876 & 2 & 8820 & \hl{0.05}\\
 & & 3 & 4900 & \hl{0.08}\\
 \hline
BH\tsb{3} & 1250 & 2 & 625 & 2.69\\
 & & 3 & 321 & \hl{0.30}\\
\hline
N\tsb{2} & 1824 & 2 & 1036 & 3.87\\
 & & 3 & 648 & 7.80\\
 & & 4 & 544 & 3.89\\
 & & 5 & 396 & \hl{0.18}\\
\hline
\hline
\end{tabular}
\label{tab:sto-3g}
\end{table}

\begin{figure}
\includegraphics[height=0.54\linewidth,width=0.95\linewidth]{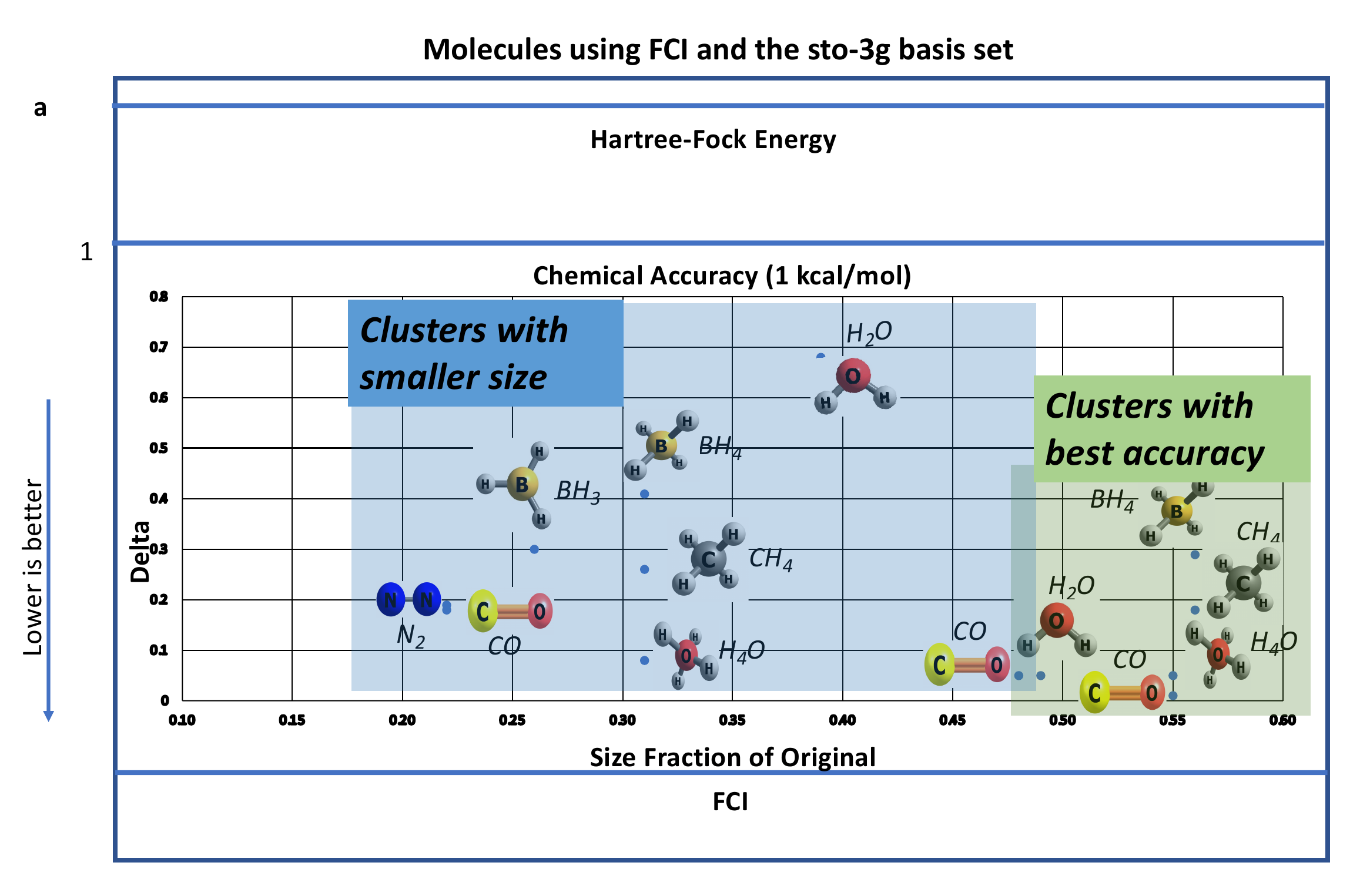}
\includegraphics[height=0.54\linewidth,width=0.95\linewidth]{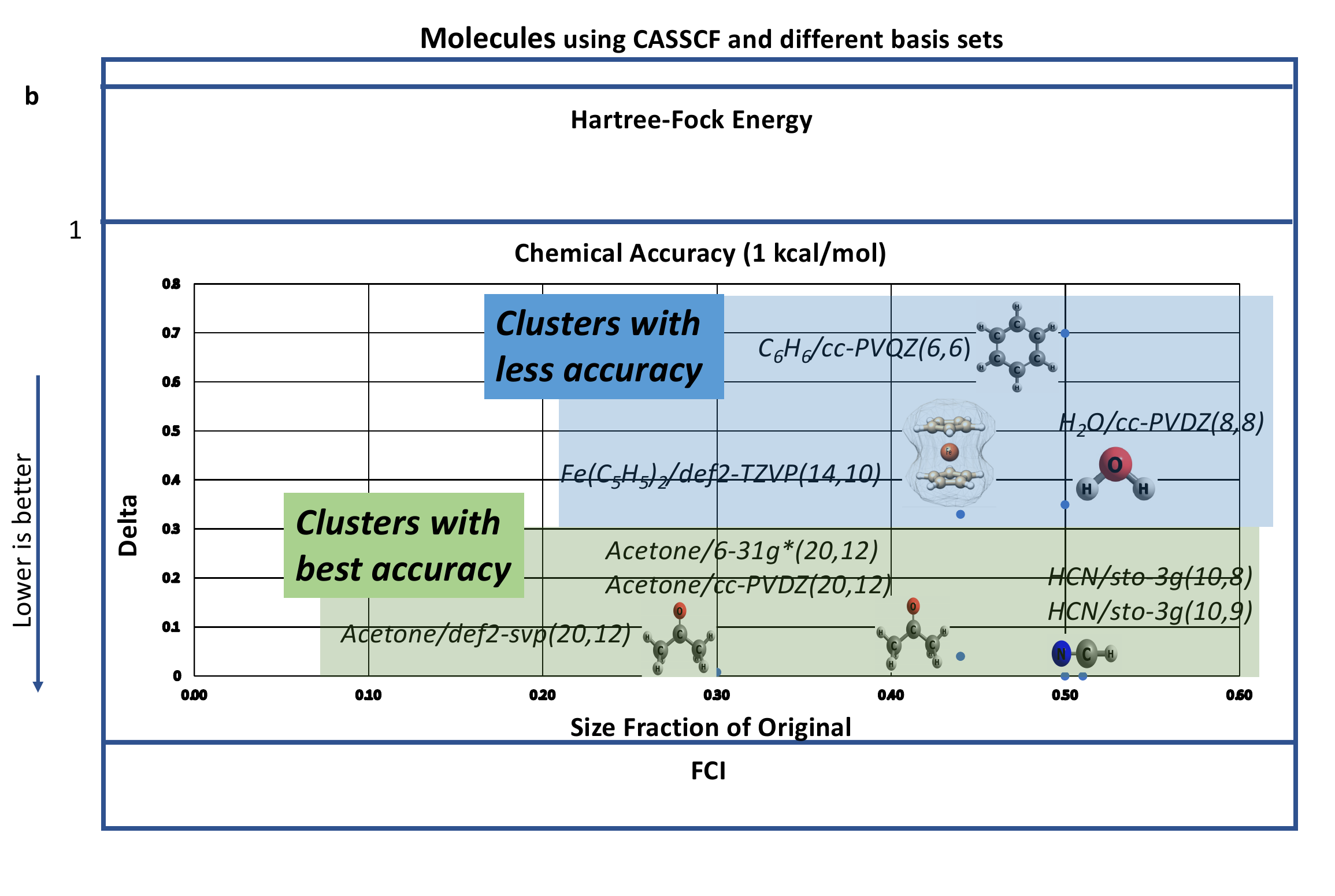}
\caption{Comparison of energy delta and size reduction for different molecules. \textbf{a} Molecule clusters using FCI and the sto-3g minimal basis set are shown. Only molecule clusters that result in energy deltas within chemical accuracy are shown. The x-axis is the size fraction from the original and the y-axis is the energy delta relative to FCI in kcal/mol. Those highlighted in light-green are the best results from an accuracy perspective, while those highlighted in light-blue have smaller size.
\textbf{b} Molecule clusters using CASSCF and different basis sets are shown. Only molecule clusters within chemical accuracy are shown. Those highlighted in light-green are the best results from an accuracy perspective, while those highlighted in light-blue have less accurate results.}
\label{fig:size_delta}
\end{figure}

\begin{table}
\centering
\caption{Approximate ground and excited states energies for H\tsb{2}O in the sto-3g basis.}
\medskip
\begin{tabular}{ccc}
\hline
Excited & $\Delta_{FCI-CL65}$ & $\Delta_{FCI-CL52}$\\
State & (kcal/mol) & (kcal/mol)\\
\hline
0 & \hl{0.05} & \hl{0.68}\\
1 & \hl{0.06} & \hl{0.19}\\
2 & \hl{0.09} & \hl{0.49}\\
3 & \hl{0.02} & \hl{0.51}\\
4 & \hl{0.04} & 33.15\\
5 & \hl{0.13} & 17.07\\
\hline
\end{tabular}
\label{tab:h2o_estates}
\end{table}

\subsection{Molecules using FCI and the sto-3g minimal basis set.}
Quantum chemists are predominantly interested in ground state energies. Molecular systems used to create FCI matrices in the sto-3g minimal basis set for further evaluation of the performance of \emph{Quantum Community Detection}
are shown in Fig.~\ref{fig:size_delta}\textbf{a}, with results in Table~\ref{tab:sto-3g}. Refer to the SI for additional data. The reduced size matrices with energy deltas within chemical accuracy are highlighted in Table~\ref{tab:sto-3g}. 
H\tsb{2}O, CO, CH\tsb{4}, BH\tsb{4}\tsp{-}, and H\tsb{4}O\tsp{2+} show a similar pattern for the \emph{k-clustering} that gave the best low energy result. The \emph{2-clustering} produced a relevant cluster that gave good initial results (within chemical accuracy, i.e. $<$ 1 kcal/mol), followed by a \emph{3-} or \emph{4-clustering} cluster that reduced the matrix size further at slightly lower accuracy. BH\tsb{3} and N\tsb{2} show a different pattern. A \emph{k-clustering} with $k >$ 2 produced a cluster with greater reduction in size and still within chemical accuracy. Figure~\ref{fig:size_delta}\textbf{a} shows the same molecules in comparison based on the size fraction of the original vs. the delta relative to FCI. Only the reduced molecule clusters resulting in energies within chemical accuracy are shown. Those highlighted in light-green from \emph{2-clustering} produced the most accurate results with $\sim$ 50\% reduction in size. Those highlighted in light-blue, from clusterings where $k >$ 2 produced clusters of smaller size, though still within chemical accuracy.

An excited state of a molecule is any electronic quantum state of the system that has a higher energy than the ground state. Table~\ref{tab:h2o_estates} shows the energy deltas for the H\tsb{2}O ground state and next five excited states for the reduced matrices based on the communities of size 65 (CL65) and 52 (CL52). Refer to the SI for additional data. Chemically accurate energies for five excited states are shown for CL65 and only the first three for CL52. This indicates that the lower accuracy CL52 does not contain all the SDs necessary for higher excited state calculations.

\begin{table}
\centering
\caption{Molecules using CASSCF and different basis sets.}
\medskip
\begin{tabular}{llcccc}
\hline
 & & & & Reduced & $\Delta_{FCI-CL}$\\
Molecule (e,o) & Basis Set & Size & k & Size & (kcal/mol)\\
\hline
\hline
HCN (10,8), acetonitrile & sto-3g & 1576 & 2 & 792 & \hl{0.000013}\\
\hline
HCN (10,9), acetonitrile & sto-3g & 8036 & 2 & 4076 & \hl{0.000054}\\
\hline
HCN (10,9), acetonitrile & {6-31G*} & 4076 & 2 & 2110 & 5.15\\
\hline
H\tsb{2}O (8,8), water & cc-PVDZ & 1234 & 2 & 617 & \hl{0.35}\\
\hline
(CH\tsb{3})\tsb{2}CO (20,12), acetone & def2-SVP & 2186 & 2 & 1027 & \hl{0.07}\\
 & & & 3 & 661 & \hl{0.0082}\\
\hline
(CH\tsb{3})\tsb{2}CO (20,12), acetone & {6-31G*} & 1098 & 2 & 513 & \hl{0.08}\\
 & & & 3 & 304 & \hl{0.04}\\
 & & & 4 & 304 & \hl{0.08}\\
\hline
(CH\tsb{3})\tsb{2}CO (20,12), acetone & cc-PVDZ & 1098  & 2 & 481 & \hl{0.04}\\
 & & & 3 & 303 & \hl{0.05}\\
 & & & 4 & 208 & \hl{0.08}\\
\hline
C\tsb{6}H\tsb{6} (6,6), benzene & cc-PVQZ & 104 & 2 & 52 & \hl{0.70}\\
 & & & 3 & 28 & \hl{0.72}\\
\hline
C\tsb{8}H\tsb{10}N\tsb{4}O\tsb{2} (12,10), caffeine & sto-3g & 7056 & 2 & 3920 & 24.36\\
\hline
Fe(C\tsb{5}H\tsb{5})\tsb{2} (14,10), ferrocene & def2-TZVP & 3632 & 2 & 1800 & \hl{0.68} \\
 & & & 3 & 1592 & \hl{0.33}\\
 & & & 4 & 1052 & 1.42\\
\hline
\hline
\end{tabular}
\label{tab:casscf}
\end{table}

\subsection{Molecules using the CASSCF method.}
The complete active space self-consistent field method (CASSCF) is a multi-configurational SCF method~\cite{Roos1980,Siegbahn1980,Siegbahn1981}. The active space consists of the electrons and orbitals that are necessary for a reliable description of the electronic structure of a molecule. CASSCF can be thought of as a truncation of FCI, which is based on a full active space within which all orbitals are considered. The total number of orbitals depends directly on the number and type of atoms in the system. 
Given a number of electrons (e) distributed across a number of orbitals (o), a Hamiltonian consisting of a linear combination of SDs or configuration state functions (CSFs) is obtained. 
Careful design of the active space requires chemical intuition based on the molecule of interest \cite{Stein2016,Keller2015}. This method is useful when HF and DFT are not adequate (frequently the case of static electronic correlations), and FCI is computationally intractable. \emph{Quantum Community Detection} could potentially be useful in the process of further truncation of the already chosen active space or to choose an optimal active space from a given initial active space approximation.

Molecules used to create the CASSCF matrices in different basis sets are shown in Fig.~\ref{fig:size_delta}\textbf{b}. For HCN and acetone, the choice of active space was standard, based on the highest occupied molecular orbital as a reference point. For benzene, ferrocene and caffeine, the choice of active space was chemically inspired, see SI.  
Acetone with different basis sets and HCN with different active spaces show the best accuracy and are highlighted in light-green. Clusters with less accuracy are seen for C\tsb{5}H\tsb{6}, Fe(C\tsb{5}H\tsb{5})\tsb{2}, and H\tsb{2}O and are highlighted in light-blue. The sizes, energies and deltas are shown in Table~\ref{tab:casscf}. 
More data is available in the SI. For most of these cases, chemical accuracy was achieved with the \emph{2-} or \emph{3-clustering} sets, as shown in Table~\ref{tab:casscf}. 
Additionally, in Table~\ref{tab:casscf}, we show two molecules, HCN (10,9) using the 6-31G* basis set and caffeine (C\tsb{8}H\tsb{10}N\tsb{4}O\tsb{2}) using the sto-3g basis set. They both 
resulted in low energy clusters greater than chemical accuracy. In the case of caffeine, expert chemical knowledge was used in choosing the original active space of 12 electrons across 10 orbitals and left no room for improvement. If the caffeine molecule with a larger initial active space of more than 12 was available, it would be expected that the optimal one with an active space of 12 would result. 

\begin{figure}
\includegraphics[width=0.90\linewidth]{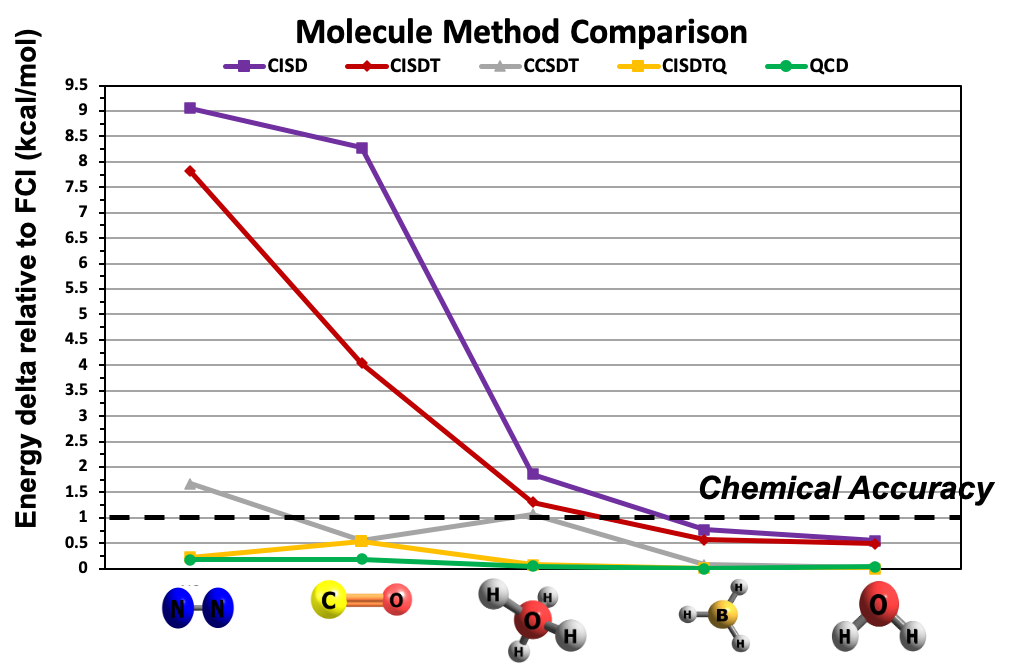} 
\caption{Comparison of methods CISD, CISDT, CISDTQ, CCSDT, and our \emph{Quantum Community Detection} (QCD) methods relative to FCI are shown for five molecules, N\tsb{2}, CO, H\tsb{4}O\tsp{2+}, BH\tsb{3}, and H\tsb{2}O. The x-axis represents each molecule. The y-axis is the energy delta relative to FCI. The lines in different colors represent the results for every method across the different molecules. CISD is shown in violet, CISDT in red, CISDTQ in gray, CCSDT in yellow, and QCD in green.
For these five molecules we can see that \emph{Quantum Community Detection} provides results that are competitive with current gold standard methods.}
\label{fig:method_compare}
\end{figure}

\subsection{Comparison with other methods.}
In Fig.~\ref{fig:method_compare}, we compare \emph{Quantum Community Detection} with commonly used current methods, CISD, CISDT, CISDTQ, and CCSDT, all relative to FCI. 
The hierarchy of approximations consists of HF $\Rightarrow$ CISD $\Rightarrow$ CISDT $\Rightarrow$ CISDTQ $\Rightarrow$ CCSDT $\Rightarrow$ FCI. The inclusion of higher excitations leads to a larger expansion and produces an answer which is closer to the exact solution. The higher configuration interaction methods are more accurate, but at the expense of an increased problem size which scales steeply ($N^6$ to $N^{10}$).
Fig.~\ref{fig:method_compare} shows the energy deltas for the five methods across the five molecules (N\tsb{2}, CO, H\tsb{4}O\tsp{2+}, BH\tsb{3}, and H\tsb{2}O) analyzed here. In this case, CISD, CISDT, and CISDTQ produce results greater than chemical accuracy. The size produced by \emph{Quantum Community Detection} compared to CISDTQ and CCSDT is smaller or similar (see SI).
In general, for this small set of molecules, we can see that \emph{Quantum Community Detection} produces results that are competitive with current gold standard methods, in line with CISDTQ and CCSDT. 

\begin{figure}
\includegraphics[width=0.70\linewidth]{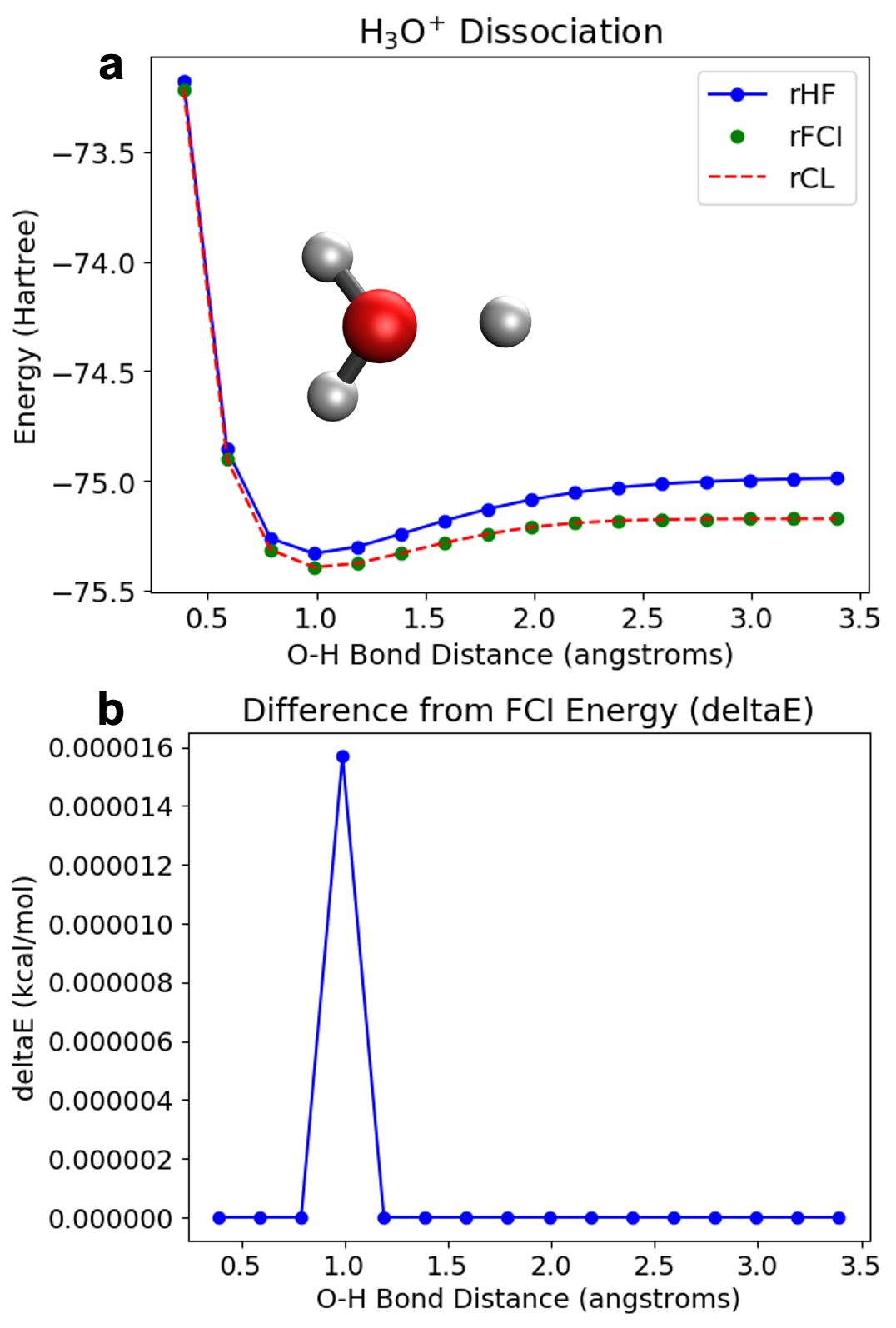}
\caption{H\tsb{3}O\tsp{+} Bond Dissociation. Showing H\tsb{3}O\tsp{+} bond dissociation using the downfolded sub-matrices discovered using \emph{Quantum Community Detection}. \textbf{a} The energies vs. the O-H bond distances are shown in the plot. The lowest energy bond distance is ~0.99 {\AA}. The FCI and cluster (CL) energies across all distances are nearly the same. \textbf{b} The energy deltas vs. the O-H bond distances are shown. Most of the energy deltas are within $10^{-10}$ Hartrees, though the lowest energy point is within $10^{-05}$ Hartrees. The reduced sub-matrix size for all bond distances is 1576 (from low energy clusters). This is a 50\% reduction from the original matrices of size 3136 SDs.}
\label{fig:h3o_dis}
\end{figure}

\subsection{H\tsb{3}O\tsp{+} bond dissociation.}
A bond dissociation energy experiment was performed for H\tsb{3}O\tsp{+} for O-H bond distances of 0.39 to 3.39 {\AA} by enforcing C$_{2V}$ molecular symmetry. We simulated a constrained potential energy surface of an oxonium ion 
(H\tsb{3}O\tsp{+}) in the sto-3g basis (FCI method), as shown in Fig.~\ref{fig:h3o_dis}\textbf{a}. Specifically, the constrained parameter was chosen as an O-H bond. Starting from the optimized geometry of H\tsb{3}O\tsp{+}, the latter was stretched/shortened by using a step of 0.2 {\AA} in a constrained scan optimization (at HF level). The overall process can be viewed as a protonation of H\tsb{2}O. The optimized geometries were used to generate FCI matrices of size 3136. The \emph{Quantum Community Detection} method resulted in reduced matrices of size 1576 for all geometries, a 50\% reduction.  Their diagonalization provided approximate ground state energies within 1.57$\times10^{-5}$ kcal/mol relative to the original FCI matrices as seen in Fig.~\ref{fig:h3o_dis}\textbf{b}. Refer to the SI for data on all the clusterings performed.

\subsection{Cluster analysis based on energy and connectivity.}
The best low energy cluster for a molecule from \emph{Quantum Community Detection} depends on the energy of the individual SDs in a cluster and the connectivity between them. Community detection determines the clusters by relying on the off-diagonal values of the Hamiltonian matrix as edge weights between SDs as the connectivity contribution.

We observe that the energy delta depends on the cluster, and it is not necessarily a monotonic function of the number of clusters within a \emph{k-clustering}. Because of this, there is an optimal $k$ after which, if $k$ is further increased, important elements from the Hamiltonian matrix are lost, and the energy delta is further increased. 

A metric to ``gauge'' the energy delta for the corresponding sub-matrix of each cluster of a particular \emph{k-clustering} can be calculated involving SD energies (diagonal elements of the Hamiltonian matrix) and the weights between SDs (off-diagonal elements of the Hamiltonian matrix). Notably, the former quantity is related to the energies of spin-orbitals composing a given SD, which historically served as a criterion for selecting active spaces for truncated methods. Therefore, an analysis of the diagonal elements of the Hamiltonian may be useful to rationalize the \emph{Quantum Community Detection} truncation.  The cluster with the lowest metric will also result in the lowest energy and smallest delta for this set. This reduces the diagonalization calculations down to one for each \emph{k-clustering}. The energy delta can be gauged with
the approximations typically used in quantum chemistry, such as perturbation theory or energy re-normalization \cite{Pastawski2001}.

\begin{figure}
\begin{center}
\includegraphics[width=0.75\linewidth]{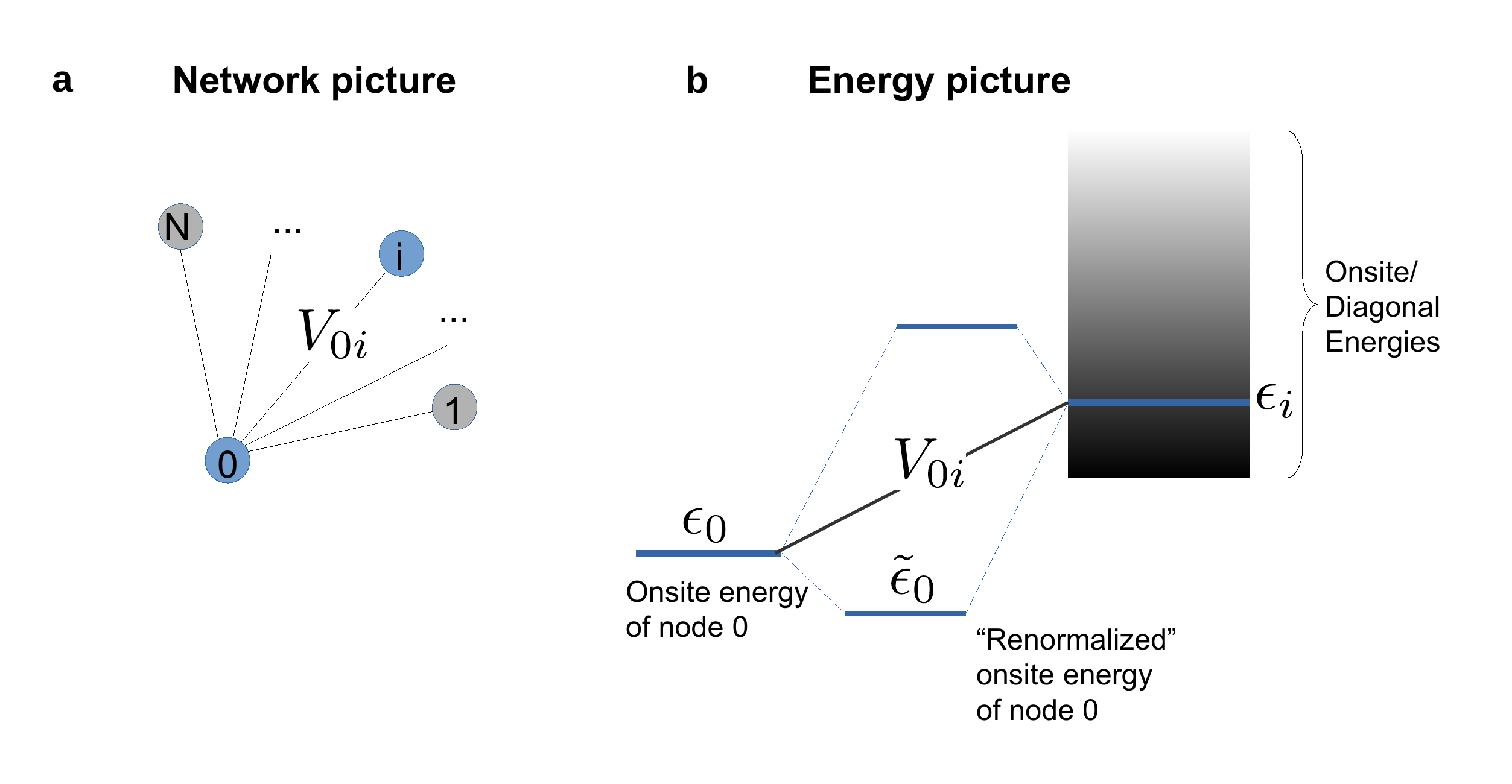}
\includegraphics[width=0.6\linewidth]{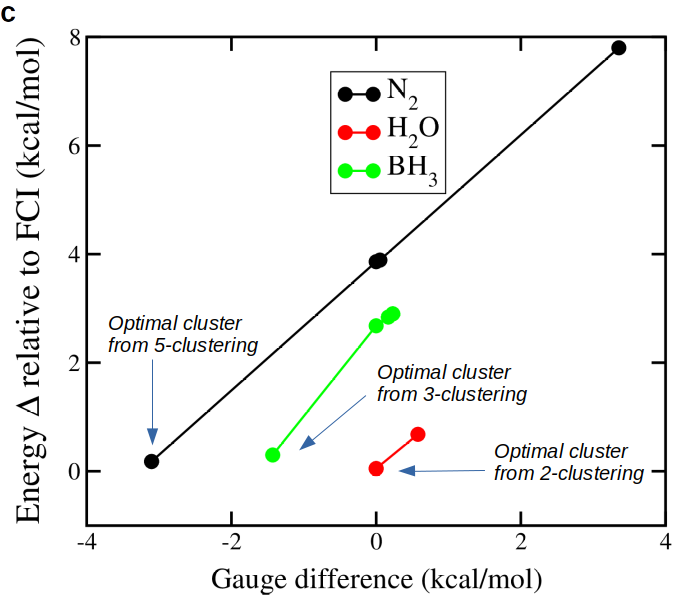}
\caption{Gauge metric description. \textbf{a} Network representation of a cluster. \textbf{b} Scheme of how the energy of node 0 (diagonal element $\epsilon_0$) gets re-normalized (corrected) to $\tilde{\epsilon}$ due to the interaction with the other nodes (SDs) via the coupling elements or weights $V_{0i}$.
\textbf{c} Gauge differences and energy deltas are shown for H\tsb{2}O, BH\tsb{3}, and N\tsb{2}. The lowest gauge metric differences and corresponding energy deltas denote the most accurate cluster. Pair-wise differences in the gauge metrics and the corresponding energy deltas have a linear relationship as shown. The most accurate cluster across all \emph{k-clusterings} is noted.}
\label{fig:gauge_metric}
\end{center}
\end{figure}

Following this same idea, we correct the energy by considering the interaction of multiple states. In Fig.~\ref{fig:gauge_metric}\textbf{a} we can see a network representation of an SD (node) interacting with many others within a network representation and within an energy representation. In Fig.~\ref{fig:gauge_metric}\textbf{b}, the energy representation, we show how the energy of node $0$ (diagonal element $\epsilon_{0}$) gets re-normalized (corrected) to $\tilde{\epsilon}_0$ due to the interaction with another node via the coupling elements or weights $V_{0i}$. The gauge metric $f$ calculated across all SDs in a cluster is shown in Eq.~\ref{gauge}.
A value of $\delta$ is added to the denominator to avoid a division by $0$. The value of $\delta$ will need to be determined for every particular problem. For the cases analized in this manuscript a value of $\delta$ = 1.0 not only shows monotonicity with $\Delta$, it is also linear. Refer to the SI for a detailed derivation.

\begin{equation}
     f = \mathrm{min}_{i} \left(\epsilon_{ii} -  \sum_k\frac{V^2_{ik}}{|\epsilon_{ii} - \epsilon_{kk}| + \delta}\right) 
\label{gauge}
\end{equation}

The metric $f$ can be used to determine the relevant cluster of a \emph{k-clustering} without having to compute $k$ diagonalizations.
Looking at the pair-wise differences in the metric $f$ values of the best clusters across multiple \emph{k-clusterings}, we see that the lowest difference determines the cluster with the best accuracy for a given molecule. This relationship between the metric $f$ differences and the energy deltas is monotonic and linear as shown in Fig.~\ref{fig:gauge_metric}\textbf{b} for H\tsb{2}O, N\tsb{2}, and BH\tsb{3}.
The most accurate cluster for H\tsb{2}O was discovered from the \emph{2-clustering} with a small energy delta of 0.05 kcal/mol. This remained the same for \emph{3-clustering} resulting in a metric $f$ difference of 0.0. For N\tsb{2}, the smallest energy delta of 0.18 kcal/mol and a low metric $f$ difference between the \emph{4-} and \emph{5-clustering} confirm that the most accurate cluster comes from the \emph{5-clustering}. Similarly, for BH\tsb{3}, the smallest energy delta of 0.30 kcal/mol and lowest metric $f$ difference between the \emph{3-} and \emph{2-clustering} indicate that the most accurate cluster is from the \emph{3-clustering}. These results are shown in Table~\ref{tab:sto-3g}. The most accurate cluster includes low energy SDs (the lower the better), which are close in energy, with high weights between the SDs (as nodes). The gauge thus provides a quick test if the inclusion of low energy SDs is sufficient for a given $k$. The gauge can also be used to indicate whether a low energy cluster will result in chemical accuracy or not (see SI).

\begin{figure}
\includegraphics[width=0.90\linewidth]{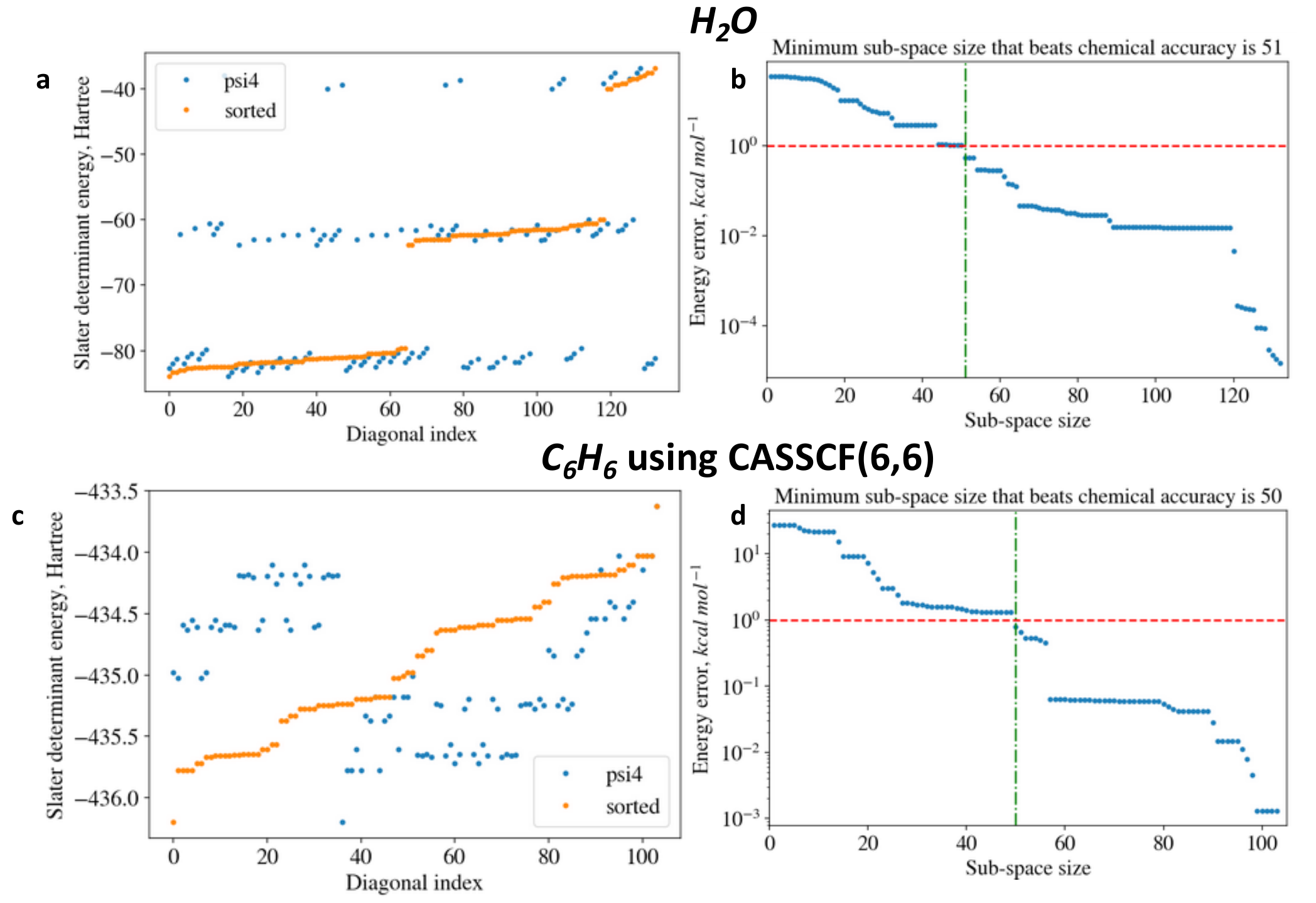}
\caption{The sorted energy approach is demonstrated for H\tsb{2}O and C\tsb{6}H\tsb{6}. \textbf{a} The SDs for H\tsb{2}O in sorted order are shown displaying distinct energy levels. \textbf{b} The minimum number of SDs required to be within chemical accuracy, 51, is shown. \textbf{c} The sorted SDs for C\tsb{6}H\tsb{6} are shown, somewhat separated in energy. \textbf{d} The minimum number of SDs within chemical accuracy, 50, is shown.}
\label{fig:sorted_energy}
\end{figure}

Could better clusters be formed from the lowest energy SDs only? Arranging the SDs in sorted order and making ``brute force'' diagonalization calculations for clusters of increasing size, showed that this approach works for some cases. In Fig.~\ref{fig:sorted_energy} we show the results for H\tsb{2}O and C\tsb{6}H\tsb{6} from a sorted energy perspective. Sorting the H\tsb{2}O SDs from low to high energy clearly shows three energy levels (see  Fig.~\ref{fig:sorted_energy}\textbf{a}). The minimum number of SDs required to be within chemical accuracy (just less than 1.0 kcal/mol) is 51 (see Fig.~\ref{fig:sorted_energy}\textbf{b}). In comparison, \emph{Quantum Community Detection} results in 65 SDs for high accuracy and 52 SDs for lower accuracy. Considering C\tsb{6}H\tsb{6}, the sorted SDs are somewhat separated in energy as seen in Fig.~\ref{fig:sorted_energy}\textbf{c}. The minimum number of SDs to be within chemical accuracy is 50, just below 1.0 kcal/mol as seen in Fig.~\ref{fig:sorted_energy}\textbf{d}. \emph{Quantum Community Detection} results in 52 SDs, with an energy delta of 0.70 kcal/mol and 28 SDs for an energy delta of 0.72 kcal/mol. This shows that the lowest energy SDs contribute a large component to the \emph{Quantum Community Detection} cluster choice, though some higher energy SDs are also present. This mix of low and high energy SDs comprise a balance producing a better result. 

\section*{Discussion}
The solution of the Schrödinger equation lies at the heart of computational chemistry and is usually accomplished by numerical diagonalization of the Hamiltonian matrix to obtain energies of the ground and excited states. Over decades, exponential hardness of the original problem was eased in practice by various many-body approximations allowing for reduction of the Hamiltonian size. Availability of quantum computing hardware provides new and unconventional means to solve traditionally difficult numerical problems. While the majority of quantum efforts have been devoted to turning exponential-to-polynomial scaling of numerical expenses, automatic reduction of the Hamiltonian size on the quantum hardware remains an important algorithmic development with practical implications. Here we have introduced a new approach for downfolding the molecular Hamiltonian matrix for electronic structure calculations using \emph{Quantum Community Detection} running on the D-Wave quantum annealer. The algorithm automatically samples the Hamiltonian space and identifies clusters allowing for accurate approximation of the lowest eigenvalue of the Hamiltonian through diagonalization that corresponds to the ground state energy of the molecule.

We have applied our \emph{Quantum Community Detection} method to a variety of different molecular systems of different sizes and chemical properties. For several cases, the resulting clusters produced ground state energies within chemical accuracy (\textasciitilde1 kcal/mol) providing size reductions of 50\% or more, and serve as good approximations to the original larger problems (i.e. calculations at full configuration interaction level). Beyond exploring chemical diversity, we have also demonstrated application to a chemical reaction/conformational dynamics example. Bond dissociation of H\tsb{3}O\tsp{+} achieved a \textasciitilde50\% reduction in matrix size while maintaining chemical accuracy.

We further have shown connections to the traditional approaches of Hamiltonian reduction via selecting molecular orbitals that provide low-energy Slater Determinants (SD).  Our analysis shows that both the energies of the SDs and the connectivity between them are important in determining the best approximate representation of a molecule within chemical accuracy for electronic structure calculations. The high accuracy clusters overall include mostly low energy SDs (the lower the better), which are close in energy. However, they may also include high energy SDs as well and the connectivity or weights between the SDs (as nodes) is typically high. The \emph{Quantum Community Detection} approach was generally able to downfold or reduce the size of the molecular Hamiltonian without chemical knowledge or intuition. For the case study of five molecules, the algorithm resulted in an efficient downfolding of Hamiltonian space toward reaching chemical accuracy, which is competitive with the current traditional approaches achieving reduction by limiting configuration interaction expansion (i.e. CISD, CISDT, CISDTQ, and CCSDT levels).

\emph{Quantum Community Detection} formulated for the D-Wave quantum annealer is a promising algorithmic development making use of emerging quantum hardware toward solving important quantum chemical problems. 
This study reports proof-of-principle concept, while practical utility of this algorithm requires further investigation.
This motivates further studies targeting exploration of 
deeper insights into relevance of quantum communities and energetics of SDs and molecular orbitals. For example, one can envision an analysis of molecular orbitals contributing to the identified communities to quantify an uncertainty in energy.
 
\section*{Methods}

\subsection{Molecular Hamiltonian matrix preparation.}
All molecular geometries were first optimized at the restricted HF level (gas) with Gaussian 09 (E01) code \cite{Frisch2009}. The FCI and CASSCF matrices were generated using the in-house modified Psi4 code \cite{Parrish2017} with  the optimized  Cartesian coordinates of atoms as input data. The matrices were generated imposing the unitary groups U(1) and SU(2) (spin and particle conservation) and point group symmetries and contain nonzero matrix elements only. If optimized by Gaussian 09, the molecule belongs to a non-Abelian point group, Psi4 automatically lowers the symmetry to one of its subgroups. The energy threshold of $10^{-8}$ $E_h$ for diagonal and of $10^{-10}$ $E_h$ for non-diagonal elements was set up. The nuclear repulsion (NR) term was added manually for clustered matrices.

\subsection{D-Wave 2000Q setup.}
The D-Wave 2000Q Ising resource at Los Alamos National Laboratory was used for this project. The solver used was \emph{DW\_2000Q\_LANL} with 2032 active qubits (out of 2048) and 5924 active couplers (out of 6016). Default parameter settings were used for all runs (with an annealing time of 20 microseconds). All runs were quantum-classical using D-Wave's \emph{qbsolv} with a sub-qubo size of 65 for the D-Wave quantum annealer.



\section*{Data Availability}
The data that support the findings of this study are available from the corresponding author upon reasonable request.

\section*{References}
\bibliography{nature_comm_dwave}


\section*{Acknowledgements}
Research presented in this article was supported by the Laboratory Directed Research and Development (LDRD) program of Los Alamos National Laboratory (LANL) under
project number 20200056DR.
This research was also supported by the U.S. Department of Energy (DOE) National Nuclear Security Administration (NNSA) Advanced Simulation and Computing (ASC) program at 
LANL. We acknowledge the ASC program at LANL for use of their Ising 2000Q quantum computing resource. 
LANL is operated by Triad National Security, LLC, for the National Nuclear Security Administration of U.S. Department of Energy (Contract No. 89233218NCA000001). Assigned: Los Alamos Unclassified Report LA-UR-20-26971.

\section*{Authorship Contributions}
S.M.M. conceived and guided the work, P.A.D. and S.M.M. designed the experiments, P.A.D. and Y.Z. generated the molecular Hamiltonians, S.M.M. ran the calculations, C.F.A.N. provided the gauge analysis, P.A.D., S.T., P.M.A., Y.Z., and S.M.M. provided other analysis. S.M.M. wrote the initial draft of the manuscript, all authors contributed to the final manuscript.

\section*{Competing Interests}
The authors declare that there are no competing interests.

\section*{Additional Information}
\begin{addendum}
 \item[Supplementary information]is available for this paper at SI.pdf.
 \item[Correspondence]and requests for materials
should be addressed to S. M. M.~(email: smm@lanl.gov).
\end{addendum}


\end{document}


\maketitle

\begin{affiliations}
 \item Computer, Computational and Statistical Sciences Division, Los Alamos National Laboratory, Los Alamos, NM
 \item Chemistry Division, Los Alamos National Laboratory, Los Alamos, NM
 \item Theoretical Division, Los Alamos National Laboratory, Los Alamos, NM
 \item Accelerator Operations and Technology Division, Los Alamos National Laboratory, Los Alamos, NM
\end{affiliations}


\section*{Supplementary Results}
Expanded tables are shown with more details including molecule point groups, FCI energies, and cluster energies.
Additional \emph{k-clustering} results are also included.

\subsection{Molecules using FCI and the sto-3g minimal basis set.}
Molecules created using FCI and the sto-3g minimal basis set are shown in Table S~\ref{tab:sto-3g-si}. 
Table S~\ref{tab:h2o_estates-si} shows the energy for the H\tsb{2}O ground state and next five excited states for the reduced matrices based on the communities of size 65 (CL65) and 52 (CL52). Chemically accurate energies for five excited states are shown for CL65 and only the first three for CL52.

\begin{table}
\centering
\caption{Molecules using the FCI method and the sto-3g minimal basis set.}
\medskip
\begin{tabular}{cccccccc}
\hline
Molecule & & $E_{FCI}$ & $E_{CL}$ & & Reduced & $\Delta_{FCI-CL}$\\
(Point Group) & Size & (Hartree) & (Hartree) & k & Size & (kcal/mol)\\
\hline
\hline
H\tsb{2}O & 133 & -75.02039100 & -75.02031690 & 2 & 65 & \hl{0.05}\\
(C\tsb{2v}) & & & -75.01930565 & 4 & 52 & \hl{0.68}\\
\hline
CO & 3648 & -111.36801449 & -111.36799244 & 2 & 2013 & \hl{0.01}\\
(C\tsb{$\infty$v} $\Rightarrow$ C\tsb{2v}) & & & -111.36793125 & 2 & 1768 & \hl{0.05}\\
 & & & -111.36793125 & 3 & 1768 & \hl{0.05}\\
 & & & -111.36016556 & 4 & 1143 & 4.92\\
 & & & -111.36770868 & 5 & 792 & \hl{0.19}\\
 \hline
CH\tsb{4} & 4076 & -39.80512049 & -39.80484039 & 2 & 2284 & \hl{0.18}\\
(T\tsb{d} $\Rightarrow$ C\tsb{2v}) & & & -39.80469968 & 3 & 1284 & \hl{0.26}\\
\hline
BH\tsb{4}\tsp{-} & 4076 & -26.61685162 & -26.61639066 & 2 & 2284 & \hl{0.29}\\
(T\tsb{d} $\Rightarrow$ C\tsb{2v}) & & & -26.61620071 & 3 & 1284 & \hl{0.41}\\
\hline
H\tsb{4}O\tsp{2+} & 15876 & -75.36527671 & -75.36519709 & 2 & 8820 & \hl{0.05}\\
(C\tsb{1}) & & & -75.36514440 & 3 & 4900 & \hl{0.08}\\
\hline
BH\tsb{3} & 1250 & -26.12145752 & -26.11717567 & 2 & 625 & 2.69\\
(C\tsb{2v}) & & & -26.12097251 & 3 & 321 & \hl{0.30}\\
\hline
N\tsb{2} & 1824 & -107.66863149 & -107.66247136 & 2 & 1036 & 3.87\\
(D\tsb{$\infty$h} $\Rightarrow$ D\tsb{2h}) & & & -107.65620751 & 3 & 648 & 7.80\\
 & & & -107.66241960 & 4 & 544 & 3.89\\
 & & & -107.66835021 & 5 & 396 & \hl{0.18}\\
 & & & -107.66835021 & 6 & 396 & \hl{0.18}\\
\hline
\hline
\end{tabular}
\label{tab:sto-3g-si}
\end{table}

\begin{table}
\centering
\caption{Excited States for H\tsb{2}O.}
\medskip
\begin{tabular}{cccccc}
\hline
Excited & $E_{FCI}$ & $E_{CL65}$ & $\Delta_{FCI-CL65}$ & $E_{CL52}$ & $\Delta_{FCI-CL52}$\\
State & (Hartree) & (Hartree) & (kcal/mol) & (Hartree) & (kcal/mol)\\
\hline
0 & -75.02039100 & -75.02031690 & \hl{0.05} & -75.01930565 & \hl{0.68}\\
1 & -74.53743573 & -74.53734050 & \hl{0.06} & -74.53713325 & \hl{0.19}\\
2 & -74.43197396 & -74.43183640 & \hl{0.09} & -74.43118524 & \hl{0.49}\\
3 & -74.31751274 & -74.31748072 & \hl{0.02} & -74.31670398 & \hl{0.51}\\
4 & -74.07950770 & -74.07943762 & \hl{0.04} & -74.02668361 & 33.15\\
5 & -74.00260411 & -74.00238742 & \hl{0.13} & -73.97540263 & 17.07\\
\hline
\end{tabular}
\label{tab:h2o_estates-si}
\end{table}

\subsection{Molecules using the CASSCF method.} 
\begin{figure}
\centering
\includegraphics[height=1.0\linewidth,width=0.60\linewidth]{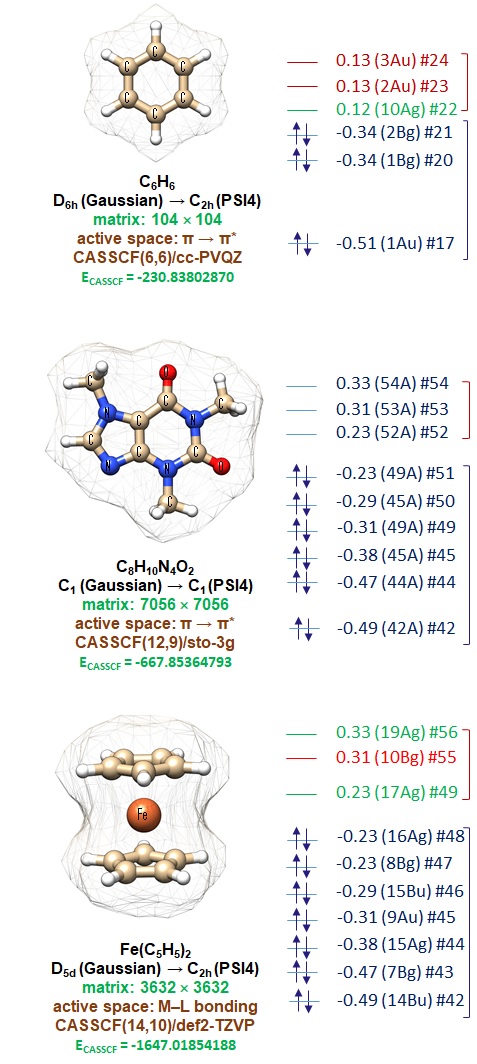} 
\caption{Chemically-inspired choice of active space for CASSCF calculations of benzene (top), caffeine (middle) and ferrocene (bottom). For each numbered molecular orbital, energy (in Hartree) and irreducible representation is shown.}
\label{fig:casscf_choice}
\end{figure}

Table S~\ref{tab:casscf-si} shows extended data for molecules using the CASSCF method in addition to what is shown in the main manuscript. The choice of active space for the benzene, caffeine, and ferrocene molecules is shown in Figure S \ref{fig:casscf_choice}.

\subsection{Molecules using other truncation methods and extended basis sets.}
Table S~\ref{tab:cis-si} shows molecules with different truncation methods and extended basis sets. Results with energy deltas within chemical accuracy are highlighted. Molecule matrices of SDs created using the CIS truncation (CI + single excitations) are chemically known to have the first determinant in the matrix (the Hartree-Fock determinant) result as the lowest eigenvalue. This means there is no correlation between the first SD and any of the other SDs. This community of 1 SD is discovered at $k=5$ for the 2 benzene examples in Table S~\ref{tab:cis-si}. The delta remains the same for $k=2$ to $k=5$ since this 1 SD is present in all.


\begin{table}
\centering
\caption{Molecules using CIS and extended basis sets.}
\medskip
\begin{tabular}{ccccccc}
\hline
Molecule & & & $E_{CL}$ & & Reduced & $\Delta_{FCI-CL}$\\
(Point Group) & Basis Set & Size & (Hartree) & k & Size & (kcal/mol)\\
\hline
\hline
C\tsb{6}H\tsb{6} & {6-31G*} & 591 & -230.70313698 & 2 & 191 & \hl{4.99e-10}\\
(D\tsb{6h}) & & & & 3 & 115 & \hl{4.99e-10}\\
 & & & & 4 & 96 & \hl{4.99e-10}\\
 & & & & 5 & 1 & \hl{4.99e-10}\\
\hline
C\tsb{6}H\tsb{6} & cc-PVDZ & 651 & -230.72234960 & 2 & 207 & \hl{5.35e-10}\\
(D\tsb{6h}) & & & & 3 & 104 & \hl{5.35e-10}\\
 & & & & 5 & 1 & \hl{5.35e-10}\\
 \hline
\hline
\end{tabular}
\label{tab:cis-si}
\end{table}

\begin{table}
\centering
\caption{Molecules using the CASSCF method.}
\medskip
\begin{tabular}{ccccccc}
\hline
Molecule & & & $E_{CL}$ & & Reduced & $\Delta_{FCI-CL}$\\
(Point Group) & Basis Set & Size & (Hartree) & k & Size & (kcal/mol)\\
\hline
\hline
HCN (10,8) & sto-3g & 1576 & -91.82114958 & 2 & 792 & \hl{0.000013}\\
(C\tsb{1}) & & & & & & \\
\hline
HCN (10,9) & sto-3g & 8036 & -91.84350714 & 2 & 4076 & \hl{0.000054}\\
(C\tsb{1}) & & & & & & \\
\hline
HCN (10,9) & {6-31G*} & 4076 & -93.00896883 & 2 & 2110 & 5.15\\
(C\tsb{1}) & & & & & & \\
\hline
H\tsb{2}O (8,8) & cc-PVDZ & 1234 & -76.10109262 & 2 & 617 & \hl{0.35}\\
(C\tsb{s}) & & & & & & \\
\hline
(CH\tsb{3})\tsb{2}CO (20,12) & def2-svp & 2186 & -191.89079792 & 2 & 1027 & \hl{0.07}\\
(C\tsb{2v}) & & & -191.89090301 & 3 & 661 & \hl{0.0082}\\
 & & & -191.89089957 & 4 & 455 & \hl{0.01}\\
 & & & -191.89089583 & 5 & 340 & \hl{0.01}\\
 & & & -191.89079142 & 6 & 391 & \hl{0.08}\\ 
 & & & -191.89079142 & 7 & 384 & \hl{0.08}\\
 & & & -191.89089583 & 8 & 343 & \hl{0.01}\\
 & & & -191.89079067 & 9 & 209 & \hl{0.08}\\
\hline
(CH\tsb{3})\tsb{2}CO (20,12) & {6-31G*} & 1098 & -192.03451734 & 2 & 513 & \hl{0.08}\\
(C\tsb{2v}) & & & -192.03456810 & 3 & 304 & \hl{0.04}\\
 & & & -192.03451389 & 4 & 208 & \hl{0.08}\\
\hline
(CH\tsb{3})\tsb{2}CO (20,12) & cc-PVDZ & 1098 & -192.05021581 & 2 & 481 & \hl{0.04}\\
(C\tsb{2v}) & & & -192.05021210 & 3 & 303 & \hl{0.05}\\
 & & & -192.05015569 & 4 & 208 & \hl{0.08}\\
\hline
C\tsb{6}H\tsb{6} (6,6) & cc-PVQZ & 104 & -230.83691596 & 2 & 52 & \hl{0.70}\\
(D\tsb{6h}) & & & -230.83688457 & 3 & 28 & \hl{0.72}\\
& & & -230.83688457 & 4 & 28 & \hl{0.72}\\
\hline
C\tsb{8}H\tsb{10}N\tsb{4}O\tsb{2} (12,10) & sto-3g & 7056 & -667.81482548 & 2 & 3920 & 24.36 \\
(C\tsb{1}) & & & -667.81195537 & 3 & 2352 & 26.16 \\
& & & -667.81195537 & 4 & 2352 & 26.16 \\
& & & -667.80878930 & 5 & 784 & 28.15 \\
& & & -667.80878930 & 6 & 784 & 28.15  \\
\hline
Fe(C\tsb{5}H\tsb{5})\tsb{2} (14,10) & def2-TZVP & 3532 & 1647.01843406 & 2 & 1800 & \hl{0.68}\\
(D\tsb{5d}) & & & -1647.01800674 & 3 & 1592 & \hl{0.33}\\
& & & -1647.01627361 & 4 & 1052 & 1.42\\
\hline
\hline
\end{tabular}
\label{tab:casscf-si}
\end{table}


\subsection{Comparison with other methods.}
In addition to the figure in the main manuscript, Table S~\ref{tab:compare-si} shows \emph{Quantum Community Detection} (QCD) compared to other methods, currently in practice, CISD, CISDT, CISDTQ, CCSDT, and FCI
for 5 molecules. Size and energy are shown for N\tsb{2}, CO, H\tsb{4}O\tsp{2+}, BH\tsp{3}, and H\tsp{2}O. QCD is comparable in energy and size for these 5 molecules.

\begin{table}
\centering
\caption{Comparison to other methods.}
\medskip
\begin{tabular}{lccccc}
\hline
Method & N\tsb{2} & CO & H\tsb{4}O\tsp{2+} & BH\tsb{3} & H\tsb{2}O\\
 & energy/size & energy/size & energy/size & energy/size & energy/size\\
\hline
\hline
CISD & -96.35/92 & -81.18/162 & -47.68/561 & -31.08/361 & -33.64/49\\
CISDT & -97.59/396 & -85.42/790 & -48.23/3041 & -31.28/1545 & -33.70/105\\
QCD & -105.23/396 & -89.27/792 & -49.49/8820 & -31.85/2450 & -34.15/65\\
CISDTQ & -105.59/1083 & -88.92/2173 & -49.46/8251 & -31.84/3355 & -34.19/133\\
CCSDT & -103.74/396 & -88.91/2173 & -49.47/8251 & -31.77/3355 & -34.16/133\\
FCI & -105.41/1824 & -89.46/3648 & -49.54/15876 & -31.85/4900 & -34.19/133\\
\hline
\hline
\end{tabular}
\label{tab:compare-si}
\end{table}

\subsection{N\tsb{2} bond dissociation.}
\begin{figure}
\includegraphics[width=0.30\linewidth]{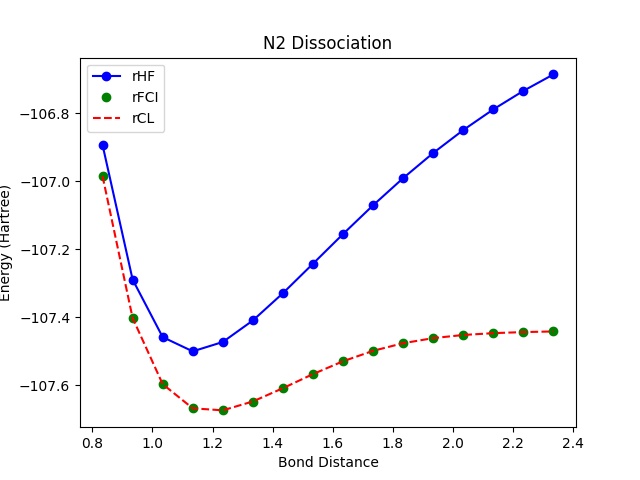} \hfill
\includegraphics[width=0.30\linewidth]{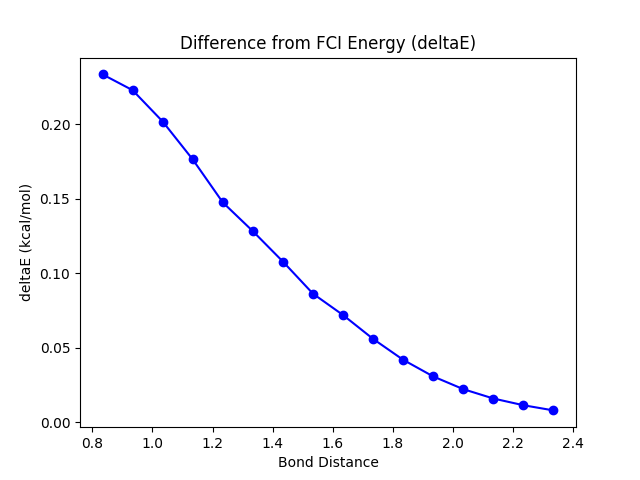} \hfill
\includegraphics[width=0.30\linewidth]{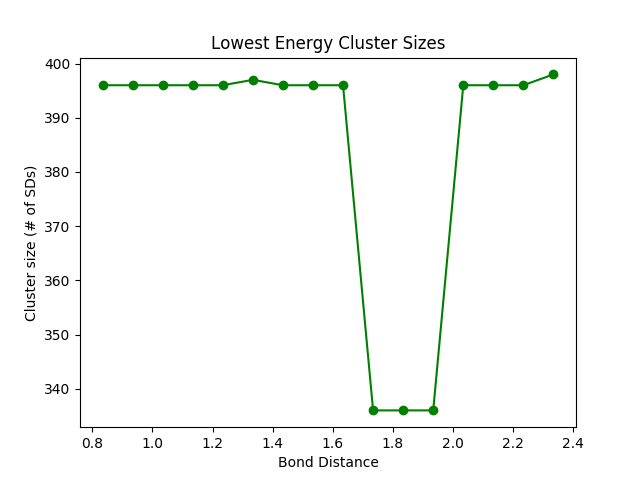}
\caption{Showing N\tsb{2} bond dissociation using the downfolded sub-matrices discovered using \emph{Quantum Community Detection}. The first picture shows energies vs. bond distances. The rFCI and rCL energies are nearly the same. The second picture shows the deltas, which decrease with increasing bond distance. And the third picture shows the reduced submatrix sizes (from low energy clusters). The original N\tsb{2} matrices of size 1824 SDs were reduced to 363-397 SDs depending on bond distance.}
\label{fig:n2_dis}
\end{figure}

A bond dissociation energy experiment was performed for $N_2$ for distances of 0.5 to 2.4 {\AA}. The ground state for each N\tsb{2} matrix was optimized using the restricted Hartree-Fock (RHF) method and the sto-3g basis set. The Hamiltonian matrices were generated using Psi4~ \cite{Parrish2017} along with the rHF and rFCI energies. Low energy communities were discovered for each bond distance using our \emph{Quantum Community Detection} approach producing energies well within chemical accuracy as seen in Fig. S~ \ref{fig:n2_dis} and Table S~\ref{tab:n2_bdis-si}. Interestingly, the $\Delta_{rFCI-rCL}$ values decrease with increasing bond distance. The downfolded sub-matrices are all similar in size ranging from 363 to 397 SDs (see Table S~ \ref{tab:n2_bdis-si}) as compared to the original 1824 SDs, up to a 25\% reduction in size.

\begin{table}
\centering
\caption{N\tsb{2} Bond Dissociation.}
\medskip
\begin{tabular}{lcccccc}
\hline
Bond & & $E_{CL}$ & & Reduced & $\Delta_{FCI-CL}$\\
Distance & Size & (Hartree) & k & Size & (kcal/mol)\\
\hline
\hline
0.83385 & 1824 & -106.98489234 & 2 & 396 & \hl{0.23}\\
\hline
0.93385 & 1824 & -107.40247938 & 2 & 396 & \hl{0.22}\\
\hline
1.03385 & 1824 & -107.59714601 & 2 & 396 & \hl{0.20}\\
\hline
1.13385 & 1824 & -107.66835022 & 2 & 396 & \hl{0.18}\\
\hline
1.23385 & 1824 & -107.67420956 & 2 & 396 & \hl{0.15}\\
\hline
1.33385 & 1824 & -107.64815357 & 2 & 397 & \hl{0.13}\\
\hline
1.43385 & 1824 & -107.60908805 & 2 & 396 & \hl{0.11}\\
\hline
1.53385 & 1824 & -107.56762396 & 2 & 396 & \hl{0.09}\\
\hline
1.63385 & 1824 & -107.52984467 & 2 & 396 & \hl{0.07}\\
\hline
1.73385 & 1824 & -107.49914796 & 2 & 336 & \hl{0.06}\\
\hline
1.83385 & 1824 & -107.47668893 & 2 & 336 & \hl{0.04}\\
\hline
1.93385 & 1824 & -107.46173680 & 2 & 396 & \hl{0.03}\\
\hline
2.03385 & 1824 & -107.45250906 & 2 & 396 & \hl{0.02}\\
\hline
2.13385 & 1824 & -107.44709122 & 2 & 396 & \hl{0.01}\\
\hline
2.23385 & 1824 & -107.44396273 & 2 & 396 & \hl{0.01}\\
\hline
2.33385 & 1824 & -107.44211517 & 2 & 398 & \hl{0.008}\\
\hline
\hline
\end{tabular}
\label{tab:n2_bdis-si}
\end{table}

\subsection{H\tsb{3}O\tsp{+} bond dissociation.}
The Tables S~\ref{tab:h3o_bdis-si}, S~\ref{tab:h3o_bdis2-si}, and S~\ref{tab:h3o_bdis3-si} show all results from all \emph{k-clusterings} that appear in the figures in the main text and more. 
A bond dissociation energy experiment was performed for H\tsb{3}O\tsp{+} for O-H bond distances of 0.39 to 3.39 {\AA}.  We simulated a constrained potential energy surface of an oxonium ion (or hydronium cation) 
(H\tsb{3}O\tsp{+}) in sto-3g basis (FCI method). Specifically, the constrained parameter was chosen as an O-H bond. Starting from the optimized geometry of H\tsb{3}O\tsp{+}, the latter was stretched/shortened by using a step of 0.2 {\AA}  in a constrained scan optimization (HF). The overall process can be viewed as a protonation of H\tsb{2}O. The optimized geometries were used to generate FCI matrices of size 3136. The \emph{Quantum Community Detection} method produced reduced matrices of size 1576 for all geometries, a 50\% reduction.  

\begin{table}
\centering
\caption{H\tsb{3}O\tsp{+} Bond Dissociation.}
\medskip
\begin{tabular}{lcccccc}
\hline
Bond & & $E_{CL}$ & & Reduced & $\Delta_{FCI-CL}$\\
Distance & Size & (Hartree) & k & Size & (kcal/mol)\\
\hline
\hline
0.39  & 3136 & -73.21424520 & 2 & 1576 & \hl{9.99e-10}\\
& & -73.21424520 & 3 & 846 & \hl{3.65e-08}\\
& & -73.21424520 & 4 & 846 & \hl{3.65e-08}\\
& & -73.21413644 & 5 & 429 & \hl{0.07}\\
& & -73.21318913 & 6 & 625 & \hl{0.66}\\
\hline
0.59 & 3136 & -74.90123035 & 2 & 1576 & \hl{9.45e-10}\\
& & -74.90123035 & 3 & 846 & \hl{9.70e-07}\\
& & -74.90123035 & 4 & 846 & \hl{9.70e-07}\\
& & -74.90108487 & 5 & 429 & \hl{0.09}\\
& & -74.90108487 & 6 & 429 & \hl{0.09}\\
\hline
0.79 & 3136 & -75.31375455 & 2 & 1576 & \hl{4.99e-10}\\
& & -75.31375455 & 3 & 846 & \hl{2.52e-07}\\
& & -75.31375455 & 4 & 846 & \hl{2.52e-07}\\
& & -75.31361997 & 5 & 429 & \hl{0.08}\\
& & -75.31361997 & 6 & 429 & \hl{0.08}\\
\hline
0.99 & 3136 & -75.39362382 & 2 & 1576 & \hl{1.57e-05}\\
& & -75.39350794 & 3 & 1225 & \hl{0.07}\\
& & -75.39350791 & 4 & 625 & \hl{0.07}\\
& & -75.39350791 & 5 & 625 & \hl{0.07}\\
& & -75.39350791 & 6 & 625 & \hl{0.07}\\
\hline
1.19 & 3136 & -75.37484922 & 2 & 1576 & \hl{1.75e-09}\\
& & -75.37474839 & 3 & 1225 & \hl{0.06}\\
& & -75.37474839 & 4 & 625 & \hl{0.06}\\
& & -75.37477659 & 5 & 925 & \hl{0.04}\\
& & -75.37474839 & 6 & 625 & \hl{0.06}\\
\hline
1.39 & 3136 & -75.32919311 & 2 & 1576 & \hl{5.35e-11}\\
& & -75.32916457 & 3 & 1276 & \hl{0.02}\\
& & -75.32910307 & 4 & 625 & \hl{0.06}\\
& & -75.32910307 & 5 & 625 & \hl{0.06}\\
& & -75.32910307 & 6 & 625 & \hl{0.06}\\
\hline
1.59 & 3136 & -75.28151591 & 2 & 1576 & \hl{9.01e-10}\\
& & -75.28143317 & 3 & 1225 & \hl{0.05}\\
& & -75.28146194 & 4 & 925 & \hl{0.03}\\
& & -75.28146194 & 5 & 925 & \hl{0.03}\\
& & -75.28143317 & 6 & 625 & \hl{0.05}\\
\hline
\hline
\end{tabular}
\label{tab:h3o_bdis-si}
\end{table}

\begin{table}
\centering
\caption{H\tsb{3}O\tsp{+} Bond Dissociation (cont'd).}
\medskip
\begin{tabular}{lcccccc}
\hline
Bond & & $E_{CL}$ & & Reduced & $\Delta_{FCI-CL}$\\
Distance & Size & (Hartree) & k & Size & (kcal/mol)\\
\hline
\hline
1.79 & 3136 & -75.24069954 & 2 & 1576 & \hl{8.92e-10}\\
& & -75.24062156 & 3 & 1225 & \hl{0.05}\\
& & -75.24065014 & 4 & 925 & \hl{0.03}\\
& & -75.24065014 & 5 & 925 & \hl{0.03}\\
& & -75.24062156 & 6 & 625 & \hl{0.05}\\
\hline
1.99 & 3136 & -75.21042253 & 2 & 1576 & \hl{5.35e-11}\\
& & -75.21034748 & 3 & 1225 & \hl{0.05}\\
& & -75.21037542 & 4 & 925 & \hl{0.03}\\
& & -75.21037542 & 5 & 825 & \hl{0.03}\\
& & -75.21034748 & 6 & 625 & \hl{0.05}\\
\hline
2.19 & 3136 & -75.19122570 & 2 & 1576 & \hl{2.32e-10}\\
& & -75.19119874 & 3 & 1276 & \hl{0.02}\\
& & -75.19117922 & 4 & 925 & \hl{0.03}\\
& & -75.19116894 & 5 & 890 & \hl{0.03}\\
& & -75.19115215 & 6 & 625 & \hl{0.05}\\
\hline
2.39 & 3136 & -75.18083769 & 2 & 1576 & \hl{4.46e-11}\\
& & -75.18081149 & 3 & 1276 & \hl{0.02}\\
& & -75.18079089 & 4 & 925 & \hl{0.03}\\
& & -75.18079089 & 5 & 925 & \hl{0.03}\\
& & -75.18076457 & 6 & 625 & \hl{0.04}\\
\hline
2.59 & 3136 & 75.17583704 & 2 & 1576 & \hl{6.06e-10}\\
& & -75.17576383 & 3 & 1077 & \hl{0.04}\\
& & -75.17576399 & 4 & 781 & \hl{0.04}\\
& & -75.17576383 & 5 & 625 & \hl{0.04}\\
& & -75.17576383 & 6 & 625 & \hl{0.04}\\
\hline
2.79 & 3136 & -75.17359540 & 2 & 1576 & \hl{1.25e-10}\\
& & -75.17359479 & 3 & 1220 & \hl{0.00039}\\
& & -75.17352726 & 4 & 897 & \hl{0.04}\\
& & -75.17352193 & 5 & 625 & \hl{0.05}\\
& & -75.17352193 & 6 & 625 & \hl{0.05}\\
\hline
2.99 & 3136 & -75.17262462 & 2 & 1576 & \hl{3.57e-10}\\
& & -75.17255090 & 3 & 1225 & \hl{0.05}\\
& & -75.17262388 & 4 & 1020 & \hl{0.00046}\\
& & -75.17255090 & 5 & 625 & \hl{0.05}\\
& & -75.17255090 & 6 & 625 & \hl{0.05}\\
\hline
\hline
\end{tabular}
\label{tab:h3o_bdis2-si}
\end{table}

\begin{table}
\centering
\caption{H\tsb{3}O\tsp{+} Bond Dissociation (cont'd).}
\medskip
\begin{tabular}{lcccccc}
\hline
Bond & & $E_{CL}$ & & Reduced & $\Delta_{FCI-CL}$\\
Distance & Size & (Hartree) & k & Size & (kcal/mol)\\
\hline
\hline
3.19 & 3136 & -75.17220417 & 2 & 1576 & \hl{6.06e-10}\\
& & -75.17180634 & 3 & 1184 & \hl{0.25}\\
& & -75.17189117 & 4 & 864 & \hl{0.20}\\
& & -75.16958650 & 5 & 760 & 1.64\\
& & -75.16947167 & 6 & 584 & 1.71\\
\hline
3.39 & 3136 & 75.17202395 & 2 & 1576 & \hl{4.19e-10}\\
& & -75.17069160 & 3 & 1176 & \hl{0.84}\\
& & -75.17069162 & 4 & 792 & \hl{0.84}\\
& & -75.17069162 & 5 & 792 & \hl{0.84}\\
& & -75.17069160 & 6 & 584 & \hl{0.84}\\
\hline
\hline
\end{tabular}
\label{tab:h3o_bdis3-si}
\end{table}

\clearpage
\subsection{Derivation for the gauge metric.}

In order to derive an expression for a metric function that could be used to gauge the error denoted by ($\Delta$) in the main manuscript, lets consider a Hamiltonian matrix $H$ composed of only two interacting states\footnote{Note that in the graph picture the interacting states will compose the nodes of a graph}. That is, we suppose that if state $i$ and $j$ are interacting through $V_{ij}$, the result of their interaction will not be affected by a third state. We have, hence, a collection of so called ``two-level systems'' represented by the following Hamiltonian matrices: 
 \begin{align}
      \left\{  
          \begin{bmatrix}
           \epsilon_{ii}  & V_{ij} \\
           V_{ij}  &    \epsilon_{jj}        
          \end{bmatrix} 
          \right\}_{i\neq j}
  \end{align}

Solving for the eigenvalues we get: 
%
%
%
%
%
%
%
\begin{equation}
\begin{split}
     \epsilon = \frac{1}{2}(\epsilon_{ii} + \epsilon_{jj}) \pm \frac{1}{2}\sqrt{(\epsilon_{ii} - \epsilon_{jj})^2 + 4V^2_{ij}}
\end{split}
\label{evals}
\end{equation}

If we expand this expression up to second order in $V_{ij}$ we get:

\begin{equation}
\begin{split}
     \epsilon \approx \frac{1}{2}(\epsilon_{ii} + \epsilon_{jj}) \pm \left( \frac{1}{2}|\epsilon_{ii} - \epsilon_{jj}|
     + \frac{V^2_{ij}}{|\epsilon_{ii} - \epsilon_{jj} |} \right)
\end{split}
\end{equation}
which leads to the following corrections to the states energies:
\begin{equation}
\begin{split}
     \tilde{\epsilon_{ii}} = \epsilon_{ii} \pm  \frac{V^2_{ij}}{|\epsilon_{ii} - \epsilon_{jj} |} \\
     \tilde{\epsilon_{jj}} = \epsilon_{jj} \pm  \frac{V^2_{ij}}{|\epsilon_{ii} - \epsilon_{jj} |} \\
\end{split}
\end{equation}
and 
\begin{equation}
\begin{split}
     \tilde{\epsilon}_{min} = \mathrm{min}\left(\epsilon_{ii} -  \frac{V^2_{ij}}{|\epsilon_{ii} - \epsilon_{jj} |},\epsilon_{jj} -  \frac{V^2_{ij}}{|\epsilon_{ii} - \epsilon_{jj} |}\right) 
\end{split}
\end{equation}
This could be extended to all the elements and further corrections by the rest of the stated to obtain the following general expression:
\begin{equation}
     \tilde{\epsilon}_{min} = \mathrm{min}_{i} \left(\epsilon_{ii} -  \sum_k\frac{V^2_{ik}}{|\epsilon_{ii} - \epsilon_{kk} |}\right) 
\label{emin}
\end{equation}
This expression can hence be used to ``gauge'' for the lowest eigenvalue and by extension, to gauge for the error $\Delta$. Note that this is similar to the Gershgorin method to have a lower/upper bound for the eigenvalues, however, with a different expression for the ``discs.''
To avoid the problem of dividing by 0 when energies are similar, a tunable parameter $\delta > 0$ was added to the denominator leading to the following metric function $f$:
\begin{equation}
     f = \mathrm{min}_{i} \left(\epsilon_{ii} -  \sum_k\frac{V^2_{ik}}{|\epsilon_{ii} - \epsilon_{kk} + \delta|}\right) 
\label{gauge}
\end{equation}
If we want an expression without having to introduce a tunable parameter, one can rewrite Eq.~\ref{evals} as follows: 
\begin{equation}
\begin{split}
     \epsilon = \epsilon_{ii} + \frac{1}{2}(\epsilon_{jj} - \epsilon_{ii}) \pm \frac{1}{2}\sqrt{(\epsilon_{jj} - \epsilon_{ii})^2 + 4V^2_{ij}}
\end{split}
\end{equation}
which gives us a correction for $\epsilon_{ii}$ provided we pick the right sign on the third term. Let's for now consider the following expression as the ``correct'' correction of $\epsilon_{ii}$:
\begin{equation}
\begin{split}
     \tilde{\epsilon}_{ii} = \epsilon_{ii} + \frac{1}{2}(\epsilon_{jj} - \epsilon_{ii}) - \mathrm{sign}(\epsilon_{jj} - \epsilon_{ii}) \frac{1}{2}\sqrt{(\epsilon_{jj} - \epsilon_{ii})^2 + 4V^2_{ij}}
\end{split}
\end{equation}
In this case, if $\epsilon_{jj} - \epsilon_{ii} > 0$, the expression becomes: 
\begin{equation}
\begin{split}
     \tilde{\epsilon}_{ii} = \epsilon_{ii} + \frac{1}{2}(\epsilon_{jj} - \epsilon_{ii}) -  \frac{1}{2}\sqrt{(\epsilon_{jj} - \epsilon_{ii})^2 + 4V^2_{ij}}
\end{split}
\end{equation}
which tends to $\epsilon_{ii}$ when $V_{ij}$ tends to 0 since the second term is positive. 
On the other hand, when $\epsilon_{jj} - \epsilon_{ii} < 0$, the expression becomes: 
\begin{equation}
\begin{split}
     \tilde{\epsilon}_{ii} = \epsilon_{ii} + \frac{1}{2}(\epsilon_{jj} - \epsilon_{ii}) +  \frac{1}{2}\sqrt{(\epsilon_{jj} - \epsilon_{ii})^2 + 4V^2_{ij}}
\end{split}
\end{equation}
which also tends to $\epsilon_{ii}$ when $V_{ij}$ tends to 0 since the second term is negative.

We can now think about another expression for the metric gauging the error which will read as follows:
\begin{equation}
\begin{split}
     g = \mathrm{min}_{i} \left(\epsilon_{ii} + \frac{1}{2} \sum_k \left( (\epsilon_{ii} - \epsilon_{jj}) -  \mathrm{sign} \sqrt{(\epsilon_{ii} - \epsilon_{jj})^2 + 4V^2_{ij}} \right)\right)
\end{split}
\end{equation}
which has the following equivalent expression\footnote{This can be easily proven considering the cases when $\epsilon_{jj} > \epsilon_{ii}$ and vice-versa}: 
\begin{equation}
     g = \mathrm{min}_{i} \left(\epsilon_{ii} + \frac{1}{2} \sum_k \left( |\epsilon_{ii} - \epsilon_{jj}| - \sqrt{(\epsilon_{ii} - \epsilon_{jj})^2 + 4V^2_{ij}} \right)\right)
\label{gauge2}
\end{equation}

Although Eq.~\ref{gauge2} is more rigorous since it does not possess any tunable parameter, it is also computationally more demanding since more operations needs to be performed. This is an important point to consider since metric $f$ is used as a gauge for the error and it will need to be computed for every cluster at every k-partition on a classical computer. This is hence one of the reasons for using metric $f$ instead of $g$ to gauge for the error in this work.

\begin{figure}
\centering
\includegraphics[width=10cm]{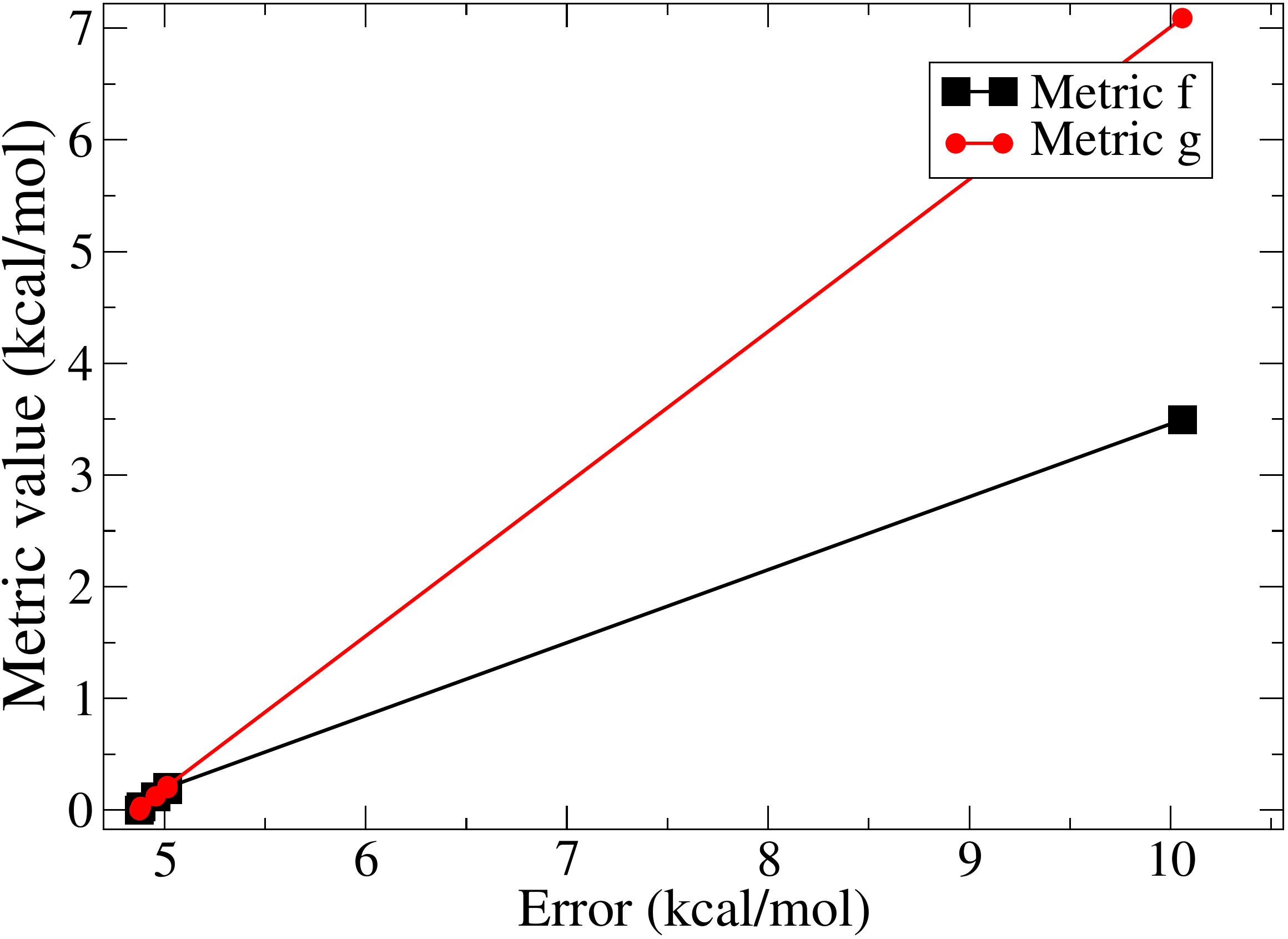} 
\caption{Comparison between metric $f$ and $g$ in their ability to gauge the error. The metric values were referred to the \emph{2-clustering} value to ease comparison. The system used here is a CO molecule FCI Hamiltonian matrix in the sto-3g basis set.}
\label{fig:gauges}
\end{figure}

Fig. S~\ref{fig:gauges} shows a comparison of both metrics on the k-clustering of molecule CO using the FCI Hamiltonian matrix with sto-3g basis set. A parameter of $\delta$ = 1 
was used for computing metric $f$. We can clearly see that both metrics agree when the errors are low whereas metric $f$ deviates from the linear trend as compared to metric $g$ when the errors are higher. For the purpose of gauging the error they are both equally suitable with the caveat that $f$ is simpler to compute. 

As was shown in the main manuscript, the metric shows both a good monotonicity and a good linearity for all the cases analyzed. It can determine the relevant cluster (without having to compute the energies for all the sub-graphs from the k-clustering). The sub-graph with the lowest gauge metric value will lead to the lowest energy (lower error). Moreover, one of the advantages of this metric is that the formula is simple enough to understand what conditions for the cluster lead to the lowest energy. 
Here we have identified three main conditions: 
\begin{itemize}
\item The lower the diagonal elements that are picked in the partition (relevant cluster), the lower will be the energy. 
\item The closer the energies of the nodes that belong to the partition that is picked, the lower will be the energy delta. 
\item The higher the coupling between nodes, the lower will be the energy delta. 
\end{itemize}
This metric, together with the results presented, can tell us when the clustering method will work better. From the expression of $f$ we can conclude that the clustering method will work better when the couplings between nodes are high as compared to the other conditions (proximity of the state energies in the relevant cluster, and the degree of how low are the individual energies of the relevant cluster).  

The gauge can also be used to indicate whether a low energy cluster will be within chemical accuracy before diagonalization. The difference between the gauge of the original matrix and the gauge of the cluster will be within chemical accuracy ($\le$ 1.6e-03 Hartrees). This is shown for a number of molecules in Table S~\ref{tab:gauge_chema-si}. H\tsb{2} molecule results were not shown previously. The others were previously seen in Table S~\ref{tab:sto-3g-si} and Table S~\ref{tab:casscf-si}. All molecule clusters that resulted in chemical accuracy were predicted correctly by using the gauge. Additionally, the gauge indicates when a cluster will not result in chemical accuracy. This could indicate that the original matrix could not be downfolded using \emph{Quantum Community Detection} or was already optimal. Further analysis is required to understand this fully. 

\begin{table}
\centering
\caption{Chemical accuracy from gauge space.}
\medskip
\begin{tabular}{lccccc}
\hline
Molecule &  & $Gauge_{FCI}$ & $Gauge_{CL}$ & Reduced & $\Delta_{FCI-CL}$\\
 & Size & (Hartrees) & (Hartrees) & Size & (Hartrees)\\
\hline
\hline
H\tsb{2} {631G*} & 8 & -1.86775811 & 0.00870209 & 4 & 0.00886926\\
H\tsb{2} cc-PVDZ & 22 & -1.86093991 & -1.85059347 & 12 & 0.01034644\\
H\tsb{2} {6-311++G**} & 54 & -1.87796529 & -1.87382138 & 29 & 0.00414391\\
H\tsb{2} aug-cc-PVQZ & 1256 & -1.88353813 & -1.87335889 & 492 & 0.01017924\\
\hline
H\tsb{2}O sto-3g & 133 & -83.91116169 & -83.91106565 & 65 & \hl{0.00009604}\\
BH\tsb{3} sto-3g & 1250 & -33.73576371 & -33.73297808 & 625 & 0.00278563\\
 & & & -33.73526023 & 321 & \hl{0.00050348}\\
N\tsb{2} sto-3g & 1824 & -130.5234671 & -130.5181132 & 1036 & 0.00535385\\
 & & & -130.5127594 & 648 & 0.0107077\\
 & & & -130.5180307 & 544 & 0.00543634\\
 & & & -130.5230616 & 396 & \hl{0.00040543}\\
\hline
HCN (8,10) sto-3g & 1576 & -115.369203 & -115.369203 & 792 & \hl{0.00000002}\\
HCN (9,10) {6-31G*} & 4076 & -117.3139781 & -117.3063941 & 2110 & 0.00758399\\
\hline
C\tsb{6}H\tsb{6} (6,6) cc-PVQZ & 104 & -436.2190515 & -436.2182797 & 52 & \hl{0.00077184}\\
C\tsb{8}H\tsb{10}N\tsb{4}O\tsb{2} (12,10) sto-3g & 7056 & -1584.331104 & -1584.316145 & 3920 & 0.01495837\\
\hline
\hline
\end{tabular}
\label{tab:gauge_chema-si}
\end{table}


\section*{References}
\bibliography{supplementary_information}